\documentclass[10pt]{article}
\usepackage{a4wide}

\usepackage{amssymb}
\usepackage{amsmath}
\usepackage[titletoc, title]{appendix}
\usepackage{graphicx} 
\usepackage[caption=false]{subfig}
\usepackage[classicReIm]{kpfonts}
\usepackage{floatrow}
\usepackage[numbers,sort&compress]{natbib}
\bibpunct[, ]{[}{]}{,}{n}{,}{,}

\usepackage{authblk}
	
\begin{document}
	

\title{An outlier-resistant indicator of anomalies among inter-laboratory comparison data with associated uncertainty}

\author{Stephen~L.~R.~Ellison}
\affil{Laboratory of the Government Chemist, Queens Road, Teddington, TW11 0LY, UK}
\date{7 August, 2018}
\maketitle
	
\begin{abstract}
A new robust pairwise statistic, the pairwise median scaled difference (MSD), is proposed for the detection of anomalous location\slash uncertainty pairs in heteroscedastic interlaboratory study data with associated uncertainties. The distribution for the IID case is presented and approximate critical values for routine use are provided. The determination of observation-specific quantiles and \textit{p}-values for heteroscedastic data, using parametric bootstrapping, is demonstrated by example. It is shown that the statistic has good power for detecting anomalies compared to a previous pairwise statistic, and offers much greater resistance to multiple outlying values. 
\end{abstract}


\noindent {\textit Submitted to:} Metrologia

\section{Introduction}
\markboth{An outlier-resistant indicator of anomalies}{\thepage}

Inter-laboratory studies are frequently used to assess the performance of methods of measurement, monitor the measurement performance of laboratories or assign reference values to materials for use in validation or calibration of future measurement systems. There are international standards and guides that govern the conduct of many such studies, particularly for test method performance and proficiency testing, one of the most common forms of laboratory performance assessment.  Most such studies involve the circulation of one or more test items to all laboratories, who then return one or more observations on each test item. 

An important part of the assessment of data in such studies is the identification of anomalous returns, either to screen data before summary statistics are calculated or to provide measures of laboratory performance. Plotting is of course an effective method of identifying outlying values where visual inspection is used, though one needs to choose the plotting method carefully when an `outlier' needs to be judged by comparison with its own reported uncertainty and not simply on the location within the data set as a whole. It is, however, usual to check that an apparently anomalous value is indeed unlikely to arise by chance. A number of statistical methods have therefore been in common use for decades for identifying anomalies in data without associated uncertainties, the most commonly used being univariate outlier tests such as Dixon's and Grubbs' tests \cite{Dixon1950}, \cite{Grubbs1950} Cochran's test for extreme variance,\cite{cochran1941} and Mandel's h and k statistics \cite{ISO5725-2} (largely used for visual inspection rather than forming an outlier check). Tests based on interquartile range are also sometimes used; points outside the traditional whisker ends on a box and whisker plot have long been understood as outliers. Outlier checks based on robust statistics can also be used; for example, a result might be regarded as suspect if outside $\hat{\mu }\pm 2\hat{\sigma }$ where $\hat{\mu }$ and $\hat{\sigma }$ are robust estimates of location and scale respectively, such as the median and scaled median absolute deviation. 

These checks are usually effective for data that is not accompanied by an estimated uncertainty and can be assumed (at least for initial inspection) to be homoscedastic. However, participants in inter-laboratory studies of measurement performance increasingly report measurement uncertainty information with their results, usually in the form of a standard uncertainty or standard error associated with the individual result. In more traditional studies, too, laboratories may report replicate observations which allow the study coordinator to assess the within-laboratory precision for each laboratory and calculate a standard error for each laboratory mean based on the dispersion of replicate observations. In either case, the coordinator then has a set of observed values each with associated uncertainty. Often, the reported or calculated uncertainties are very different, that is, the results must be considered heteroscedastic. Where that is the case, and particularly where summary statistics use the reported uncertainty, classical outlier tests are inappropriate; a comparatively extreme result accompanied by a (correctly estimated) large uncertainty would be incorrectly classed as anomalous by classical outlier tests, while a more central result with a very small standard uncertainty may be mistakenly treated as unexceptional. There is therefore good reason to seek indicators of anomalous data which take associated uncertainty into account.

In this study, we propose an indicator based on a scaled pairwise comparison to identify anomalous values. A brief motivating example is given, followed by the proposed statistic. The distribution and critical values for the heteroscedastic case are considered and critical values for that case are provided. A methodology for assessing statistical significance for the important case of heteroscedastic data is given. Finally, the performance of the proposed statistic is compared with another, related, statistic that has been proposed for this purpose.

\section{A motivating example}

\begin{figure}[h!]
	\includegraphics*[width=5.00in, keepaspectratio=true]{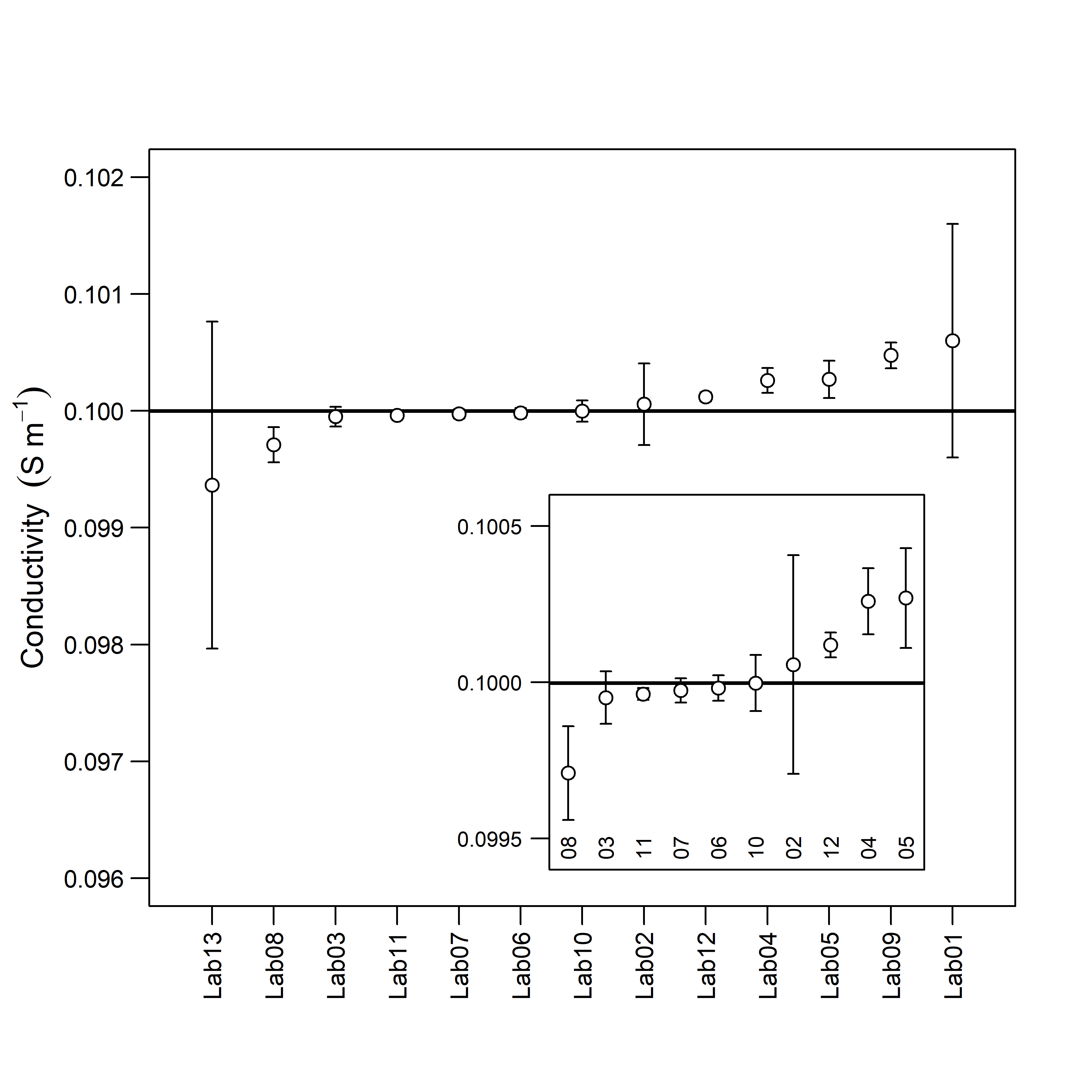}
	\caption{Results from an interlaboratory comparison in electrochemistry (specifically, conductivity of standard buffer solutions) in which laboratories reported uncertainties. Error bars show `expanded uncertainties' at $\pm2u$, broadly interpretable as estimated 95\% confidence intervals with allowance for possible systematic effects, calibration uncertainties etc. The inset shows a vertical expansion of the central laboratories, with laboratory numbers along the bottom of the inset. The horizontal line is the median of the data set.}
	\label{fig:ccqm-p22}
\end{figure}

\begin{table}
	\caption{ \label{table:table1}Conductivity measurements in an international comparison }
	{\begin{tabular}{lcc} 
			Laboratory & Conductivity & $u$\textsuperscript{a}\\ 
			~ & S~cm\textsuperscript{-1} & S~cm${}^{-1}$ $\boldsymbol{\times}$1000 \\ 
			Lab13 & 0.099365 & 0.7000 \\ 
			Lab08 & 0.099710 & 0.0750 \\ 
			Lab03 & 0.099951 & 0.0420 \\ 
			Lab11 & 0.099963 & 0.0095 \\ 
			Lab07 & 0.099974 & 0.0195 \\ 
			Lab06 & 0.099982 & 0.0205 \\ 
			Lab10 & 0.099998 & 0.0450 \\ 
			Lab02 & 0.100057 & 0.1750 \\ 
			Lab12 & 0.100120 & 0.0200 \\ 
			Lab04 & 0.100260 & 0.0530 \\ 
			Lab05 & 0.100270 & 0.0800 \\ 
			Lab09 & 0.100475 & 0.0550 \\ 
			Lab01 & 0.100600 & 0.5000 \\ 
	\end{tabular}}\\
	\textsuperscript{a}~Reported standard uncertainty in measured conductivity. Note the multiplication for readability.
	
\end{table}

Consider the data in Table \ref{table:table1}, plotted in Figure~\ref{fig:ccqm-p22}
The data are from a pilot exercise on conductivity \cite{CCQM-P22}, organised under the auspices of the CCQM, the international committee responsible for measurements of amount-of-substance. First, the plot shows very wide differences in reported uncertainty. These are largely attributable to differences in available equipment in different institutes. Second, some apparently extreme values at either end (the values are plotted in ascending order of result) appear to be outliers, some severe. Yet closer inspection shows that these values are accompanied by large confidence intervals, overlapping both the central line in the plot (here, drawn through the median of the data to aid inspection) and each other's intervals. By contrast, some of the central values with small uncertainties do not appear to be outliers at all, yet on closer inspection (inset)  it becomes apparent that these laboratories do not agree with one another within reported uncertainties, and some do not agree with the central location estimate. To complicate interpretation, agreement (or otherwise) with the central location estimate depends on the choice of estimator; changing from a median to a mean, to a weighted mean, to a weighted robust estimate (such as an MM-estimate \cite{Yohai1976}) or to a random-effects model with an additional variance term for overdispersion, changes the interpretation as to which of these laboratories is performing well and which is not. Finally, for many outlier detection methods, the presence of two or more outliers is very likely to cause some `masking', and some of the values with larger reported uncertainty in this case may be marked as outliers on the basis of their location, when in fact their reported uncertainties show their location is not unreasonable.

In such cases, it is useful to have some indicator of anomalies that a) does not depend on the location estimated from the data, b) uses the reported uncertainty information and c) is insensitive to a fairly large proportion of anomalous values,  `anomalous' being judged in terms of location compared to the associated uncertainty.

\section{A pairwise median statistic for detection of location\slash uncertainty observations}

Douglas and Steele \cite{Steele2006} have previously suggested the use of pair-difference chi-squared statistics as an indicator of anomalies or overdispersion in similar data, an approach which removes dependence on a choice of location estimator but which remains sensitive to multiple outliers. Rousseeuw \cite{Rousseew1993} has shown that a quantile-based function of all pairwise differences provides an efficient outlier-resistant estimator of scale. Inspired by this pairwise approach, and seeking a more outlier-resistant indicator of individual anomaly than a chi-squared statistic, the use of a median absolute scaled difference (MSD) is proposed as an indicator of anomalous data. This is calculated simply as follows:

\renewcommand{\theenumi}{\roman{enumi}}
\begin{enumerate}
	\item For each value/uncertainty pair $(x_{i}, u_{i})$ form the set $d_{i}$ of $j=1..n, j~\neq~i$ scaled differences as 
\begin{equation}
d_{ij} =\frac{x_{i} -x_{j} }{\sqrt{u_{i}^{2} +u_{j}^{2} } }  
\label{eqn:1}
\end{equation}
 
	\item Calculate the median absolute scaled deviations $Q_{{\rm E}_{i} }$ as 
\begin{equation}
Q_{{\rm E}_{i} } ={\rm med}\left(\left|d_{ij} \right|,\; j=1..n,\, j\ne i\right) 
\label{eqn:2}
\end{equation}
 
($Q$ is chosen here for the similarity to Rousseuw's $Q_{n}$). 
\end{enumerate}

Inspection of eq. \eqref{eqn:2} gives a qualitative idea of how $Q_{\rm E}$ will behave.  The $d_{ij}$ are formed by dividing a signed difference by the (estimated) uncertainty of the difference. If $x_{i}$ and $x_{j}$ arise from normal distributions with a common mean value and with standard deviations $u_{i}$ and $u_{j}$ respectively, we expect the distribution of $d_{ij}$ to be normal with mean 0 and standard deviation 1. The median of the absolute values of a standard normal distribution is 0.674. However, as $x_{i}$ becomes discrepant with more and more $x_{j}$, the indicator increases; values far above 0.674 indicate that the $x_{i}$ is discrepant with a majority of other data points. Because this indicator is based on the median of differences, it is not greatly affected by a small number of large differences. The indicator uses all the available uncertainty information. Finally, because the indicator is based solely on pairwise differences, it is independent of any estimated location for the data set as a whole, satisfying all three of the desirable properties listed above.

It is, however, important to have some indication of how large $Q_{\rm E}$ needs to be to indicate that an apparent anomaly is unlikely to arise by chance; that is, some guidelines or critical value that characterises a data point as unusually extreme when judged by this indicator. It is also sensible to ask how sensitive the indicator is to anomalies; that is, how powerful $Q_{\rm E}$ is as a test for anomaly in the presence of other `outlying' values. These are the subjects of the next three sections.

\section{Distribution and critical values for the IID case}
The distribution for $Q_{\rm E}$ in the IID case is clearly of limited interest when faced with highly heteroscedastic data such as that in Figure~\ref{fig:ccqm-p22}. It does, however, provide an understanding of the expected behaviour, it is of interest when the data show roughly similar uncertainties and, as we show below, it provides a rough guideline for inspection. We therefore present brief results for the IID case here, with derivation in Appendix 1. 

For the case of $n$ even, $Q_{\rm E}$ for a single observation $x_{0}$ chosen at random is the median of $n-1$ absolute differences {\textbar}$x_{1}$~--~$x_{0}${\textbar}, {\textbar}$x_{2}$~-~$x_{0}${\textbar}, {\dots},  {\textbar}$x_{n-1}$~-~$x_{0}${\textbar}. Appendix 1 shows that the cumulative distribution $F_{Q_{{\rm E}} } \left(\left|d\right|\right)$ conditional on $x_{0}$ can be written in a computationally useful and compact incomplete beta function form,
\begin{equation}
F_{Q_{{\rm E}} } \left(\left|d\right|\right)=\int \limits _{-\infty }^{\infty }I_{F_{\left|D\right|} \left(\left|d\right||x_{0} \right)} \left(r,n-r\right)\varphi (x_{0} )dx_{0}   
\label{eqn:3}
\end{equation}
 
where 
\begin{equation}
F_{\left|D\right|} \left(\left|d\right||x_{0} \right)=\Phi \left(x_{0} +\left|d_{i} \right|\sqrt{2} \right)-\Phi \left(x_{0} -\left|d_{i} \right|\sqrt{2} \right) 
\label{eqn:4}
\end{equation}
 
and $I_{p}(a, b)$ is the incomplete beta function at a value $p$ with parameters $a$ and $b$ (see Appendix 1 for definition). Integration over $x_{0}$, assumed normally distributed with mean and standard deviation identical to the true mean and standard deviation of all other observations, then provides the marginal distribution. 

The form for odd $n$, which is a median of an even number of differences, is more intricate; a form conditional on $x_{0}$ can be shown (Appendix 1 part 2) to be
\begin{equation}
F_{Q_{x} } \left(\left|d\right||x_{0} \right)=\frac{2}{B(r_{{\rm e}} ,r_{{\rm e}} )} \int \limits _{0}^{\left|d\right|}\left[F_{\left|D\right|} \left(t|x_{0} \right)\right]^{r-1} \left\{\left[1-F_{\left|D\right|} \left(t|x_{0} \right)\right]^{r} -\left[1-F\left(2\left|d\right|-t|x_{0} \right)\right]^{r} \right\}f_{\left|D\right|} \left(t\right)d(t)  
\label{eqn:5}
\end{equation}
 
\begin{figure}[h!]
	\includegraphics*[width=3.50in, keepaspectratio=true]{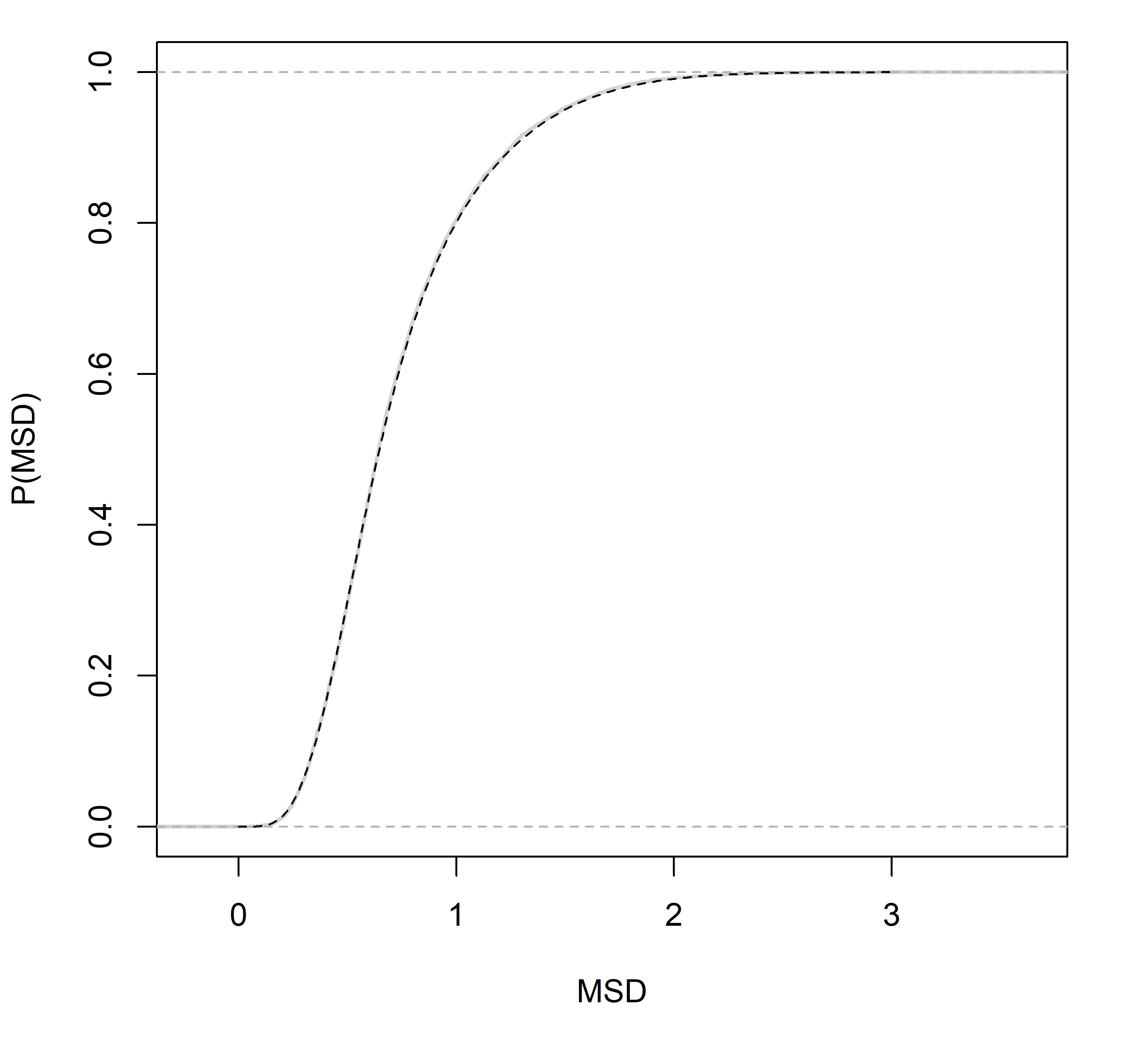}
	\caption{ The figure shows an empirical cumulative distribution (grey) for 10000 simulations of calculations of median absolute scaled difference (MSD, denoted $Q_{\rm E}$ in the text) for the first point in n=10 normally distributed values. The dashed line (black) shows the cumulative distribution function calculated by numerical integration of equation \eqref{eqn:3}.}
	\label{fig:ecdf}
\end{figure}

In addition to the integration over $t$, this also requires integration over $x_{0}$ to obtain the marginal distribution. Neither form is amenable to algebraic integration; both must be evaluated numerically. This is straightforward for the even-$n$ case (eq. \eqref{eqn:3}) up to at least $n=10^{6}$, as we can take advantage of robust functions implementations of the incomplete beta function. With care, integration is also achievable, albeit more slowly, for the odd-$n$ case of eq. \eqref{eqn:5} at least up to $n=299$, though we found that numerical rounding becomes a problem at high probabilities and higher $n$ (near 400). An example distribution function calculated using eq. \eqref{eqn:3} for $n=10$ is given as Figure~\ref{fig:ecdf}, compared with the empirical cumulative distribution from a small simulation. Given a cumulative distribution implementation, quantiles for any reasonable probability and value of $n$ can be obtained by numerical root finding. Interpolation on pre-calculated values can also provide useful estimates quickly; practical estimation for routine use is discussed in more detail in Appendix 2. 

The limitation to modest odd $n$ is not a serious problem. First, the likely applications involve smaller data sets; typical metrology comparisons involve fewer than 30 laboratories. Second, investigation of estimation errors showed that the next higher even-$n$ case provides an excellent approximation for higher odd $n$. For $n>99$ the worst-case differences in probability estimation between the $n$ and $n+1$ cases (located by grid search and subsequent maximisation of the difference) were never more than $5\times10^{-5}$, with these worst case differences occurring, unsurprisingly, at the steepest part of the cumulative distribution curve (around $p=0.3$). For probabilities of 0.95 or above the error in probability estimates due to use of the next even $n$ case were under $10^{-6}$ for all odd $n$ from 101 to 301. Quantiles for odd $n$ could also be estimated from the next even $n$ to far better numerical accuracy than normal tabulation requires; for $n\mathrm{\ge}99$, errors in quantile estimation from using the next higher even $n$ were all under $4\times10^{-5}$, and less than $10^{-6}$ for probabilities over 0.8. This is inconsequential for typical quantiles. For practical purposes, therefore, it seems safe to use equation \eqref{eqn:5} for odd $n$ up to $n=99$ and equation \eqref{eqn:3} applied to the next higher even $n$ thereafter. 

An understanding of the asymptotic distribution as $n$ grows is also of interest, both for insight and for interpolation, where an exact limiting value aids cubic or spline interpolation near the extremes and also permits interpolation to very large $n$ in tables. The cumulative distribution $\mathop{\lim }\limits_{n\to \infty } F_{Q_{{\rm E}} } \left(\left|d\right|\right)$ is derived in Appendix 1 part 3 (eq \eqref{eqn:22}). Although again requiring numerical evaluation, this does not involve a numerical integral once an implementation of the Normal distribution is available; probabilities and quantiles can be located rapidly by standard root finding methods. These are practically useful as a limiting value for interpolation to very large $n$ than can conveniently be tabulated. An interesting feature of the limiting distribution is that it is bounded to the left at $0.674/\sqrt{2}=0.477$, a feature that arises from the fact that the population median of a half-normal distribution is 0.674. This accounts for the general form of the distribution in Figure~\ref{fig:ecdf}, which is, as expected, zero at $|d|=0$, and increases sharply as it approaches 0.477. This has some interesting effects on the quantiles as $n$ increases; quantiles, especially for lower probabilities from 0.1 to 0.3, are not necessarily monotonic with increasing $n$ as the shape of the curve changes with increasing $n$, a factor to bear in mind if interpolating between values for $n$.  

\begin{figure}
	\centering
	\sidesubfloat[]{%
		\resizebox{3.5 in}{!}{%
			\includegraphics[trim={0 0 0 0.25in }, keepaspectratio=true]{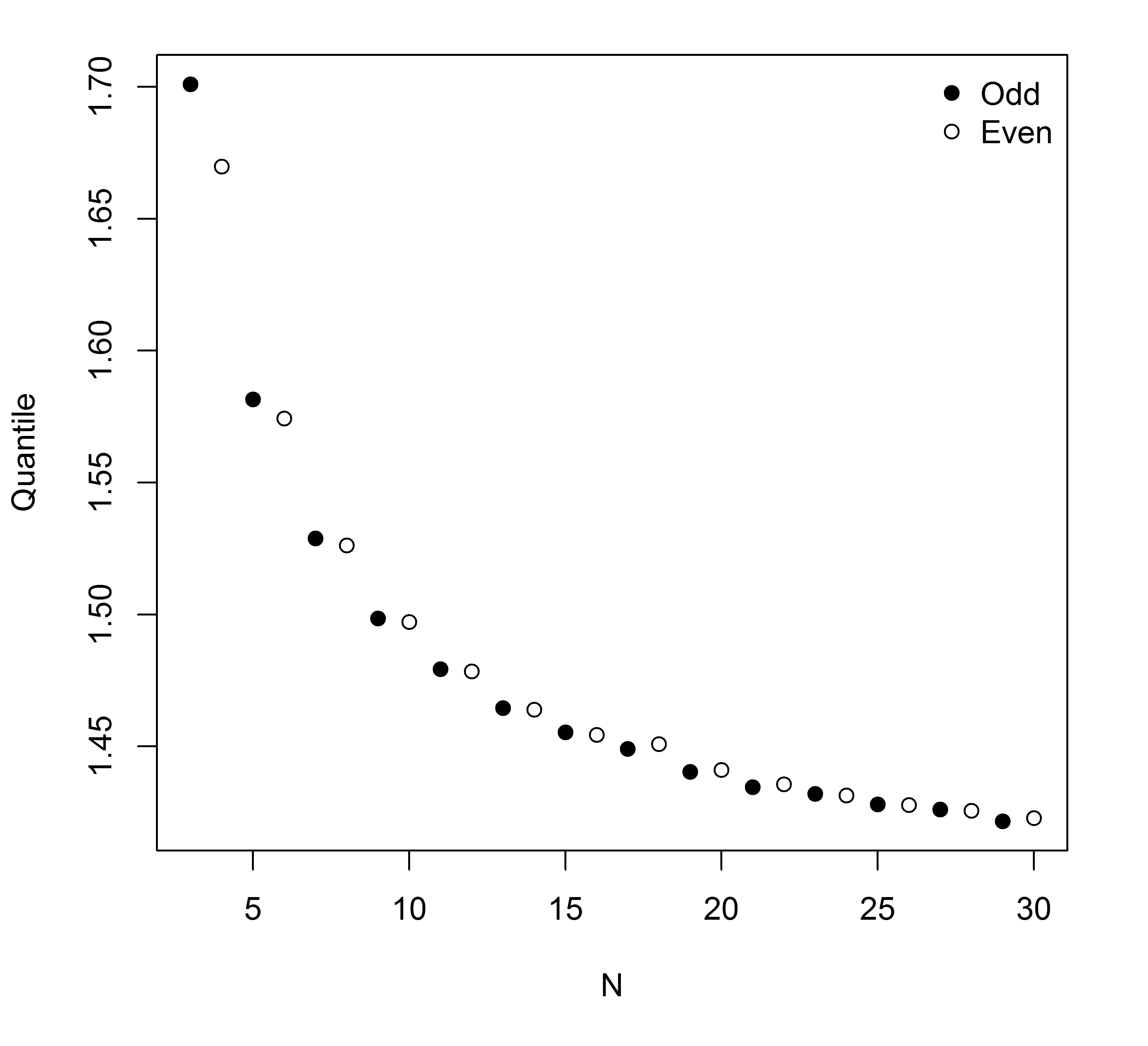} }
		\label{fig:3a}
	}
	
	\sidesubfloat[]{%
		\resizebox{3.5 in}{!}{%
			\includegraphics[trim={0 0 0 0.25in }, keepaspectratio=true]{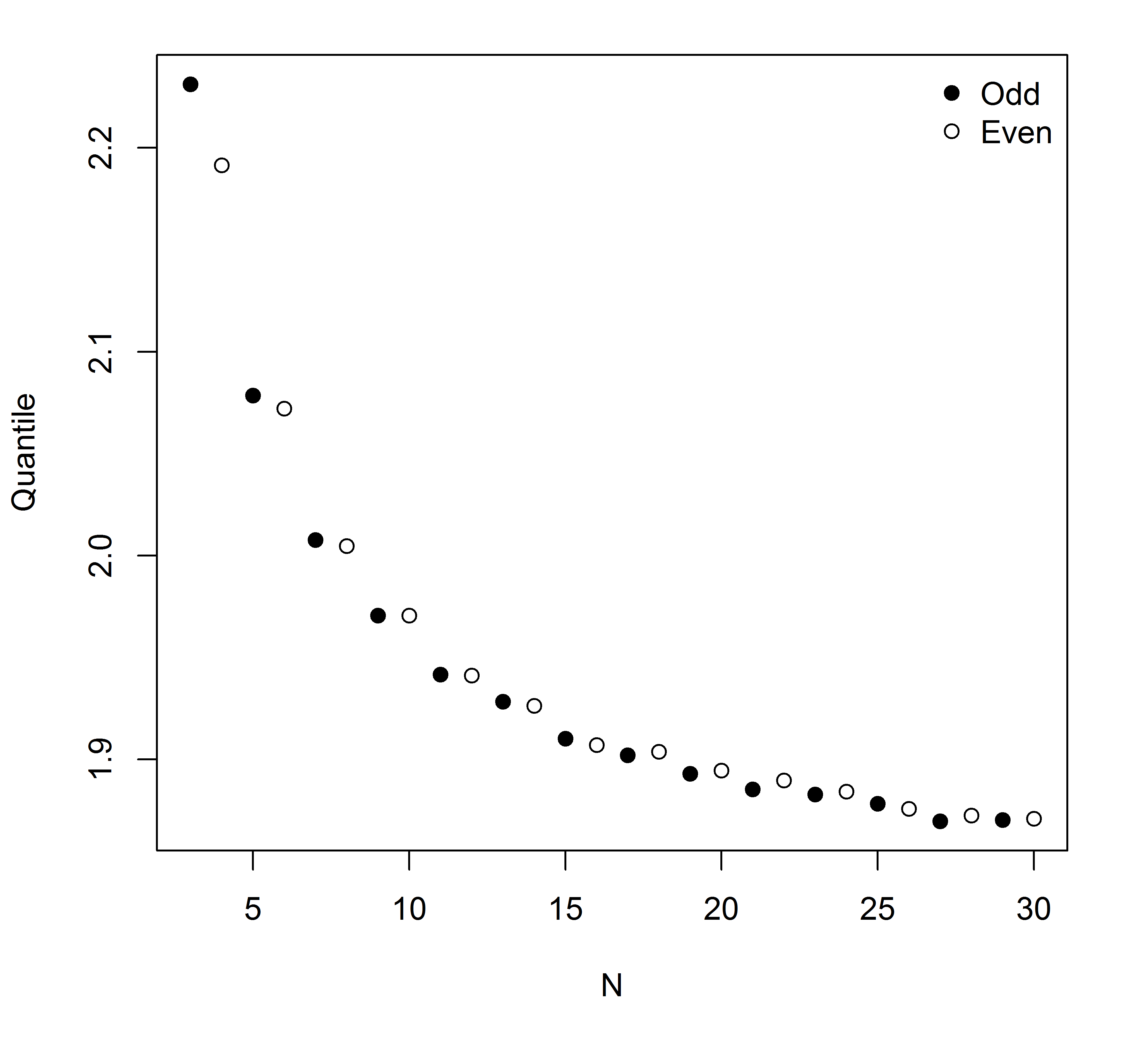} } 
		\label{fig:3b}
	}
	\caption{The figure shows upper quantiles at a) 95\% and b) 99\% calculated for each of the values of $n$ shown, using numerical integration of equations \eqref{eqn:3} and \eqref{eqn:5} as appropriate. Filled circles are for odd $n$, open circles for even $n$.}
	\label{fig:quantiles}
\end{figure}

A set of quantiles for common probabilities and values of $n$ is provided in Table \ref{table:table2}, which also includes asymptotic values in the final row.  Figure~\ref{fig:quantiles} shows 95\% and 99\% upper quantiles for $n$ up to 30. As the Figure shows clearly, the quantiles for the even $n$ case follow a different sequence to the odd-$n$ case. The table is accordingly divided into two parts, for odd $n$ and even $n$. To use the table for intermediate values of $n$, one can conveniently use the next lower value of $n$ in the relevant part of the table (odd or even).

In addition, functions for calculating arbitrary quantiles, densities and probabilities are provided in the R package metRology \cite{metRology}, with source code available at the cited URL. A table that can be used for rapid interpolation (strictly, two tables, for odd and even $n$ respectively) is also provided with this manuscript as electronic supplementary information.

\floatsetup[figure]{style=plain,subcapbesideposition=top}

\floatsetup[table]{style=plaintop, capposition=top}
\begin{table}
	\begin{tabular}{p{0.3in}p{0.5in}p{0.5in}p{0.5in}p{0.5in}p{0.5in}p{0.5in}} 
		\multicolumn{7}{p{4.5 in}}{a) $N$ even} \\ 
		$N$ & \multicolumn{6}{p{2.7in}}{Probability $p$} \\ 
		& 0.5 & 0.75 & 0.9 & 0.95 & 0.99 & 0.999 \\ 
		4 & 0.664 & 1.014 & 1.407 & 1.670 & 2.193 & 2.803 \\ 
		6 & 0.657 & 0.961 & 1.325 & 1.573 & 2.067 & 2.641 \\ 
		8 & 0.651 & 0.931 & 1.283 & 1.525 & 2.004 & 2.561 \\ 
		10 & 0.647 & 0.912 & 1.259 & 1.497 & 1.967 & 2.513 \\ 
		12 & 0.643 & 0.899 & 1.243 & 1.478 & 1.942 & 2.481 \\ 
		14 & 0.640 & 0.889 & 1.231 & 1.465 & 1.925 & 2.459 \\ 
		16 & 0.637 & 0.882 & 1.223 & 1.455 & 1.912 & 2.442 \\ 
		18 & 0.634 & 0.876 & 1.216 & 1.447 & 1.902 & 2.429 \\ 
		20 & 0.632 & 0.871 & 1.211 & 1.441 & 1.894 & 2.419 \\ 
		22 & 0.630 & 0.868 & 1.206 & 1.436 & 1.887 & 2.411 \\ 
		24 & 0.629 & 0.864 & 1.203 & 1.432 & 1.881 & 2.403 \\ 
		26 & 0.627 & 0.862 & 1.200 & 1.428 & 1.877 & 2.398 \\ 
		28 & 0.625 & 0.859 & 1.197 & 1.425 & 1.873 & 2.392 \\ 
		30 & 0.624 & 0.857 & 1.195 & 1.423 & 1.869 & 2.388 \\ 
		40 & 0.619 & 0.851 & 1.187 & 1.413 & 1.857 & 2.373 \\ 
		50 & 0.615 & 0.847 & 1.183 & 1.408 & 1.850 & 2.363 \\ 
		60 & 0.612 & 0.844 & 1.179 & 1.404 & 1.845 & 2.357 \\ 
		70 & 0.610 & 0.842 & 1.177 & 1.402 & 1.842 & 2.353 \\ 
		80 & 0.608 & 0.841 & 1.176 & 1.400 & 1.839 & 2.350 \\ 
		90 & 0.606 & 0.840 & 1.174 & 1.398 & 1.837 & 2.347 \\ 
		100 & 0.605 & 0.839 & 1.173 & 1.397 & 1.836 & 2.345 \\ 
		$\infty$ & 0.593 & 0.831 & 1.164 & 1.386 & 1.821 & 2.327 \\ 
		\multicolumn{7}{p{4.5 in}}{%
			$N$ is the total number of observations, including the observation for which $Q_{\rm E}$ is calculated. Quantiles were calculated from eq. \eqref{eqn:3} or (for the asymptotic values in the last row) eq. \eqref{eqn:22}, and are rounded to three decimal places. The probabilities are lower tail probabilities, so that quantiles are upper critical values for significance $1-p$. To use the table for intermediate values of $N$, use the next lower value of $N$ in the relevant part of the table (odd or even). See also Appendix 2 for discussion of interpolation methods for $N \gg 100$.} \\
	\end{tabular}
	\caption{Single-observation upper quantiles for median scaled absolute difference}
	\label{table:table2}
\end{table}
\begin{table}	
	\begin{tabular}{p{0.3in}p{0.5in}p{0.5in}p{0.5in}p{0.5in}p{0.5in}p{0.5in}} 
		\multicolumn{7}{p{4.5 in}}{b) $N$ odd} \\ 
		$N$ & \multicolumn{6}{p{2.7in}}{Probability $p$} \\ 
		& 0.5 & 0.75 & 0.9 & 0.95 & 0.99 & 0.999 \\ 
		3 & 0.714 & 1.055 & 1.440 & 1.702 & 2.231 & 2.850 \\ 
		5 & 0.672 & 0.972 & 1.332 & 1.581 & 2.076 & 2.652 \\ 
		7 & 0.658 & 0.936 & 1.286 & 1.528 & 2.008 & 2.565 \\ 
		9 & 0.651 & 0.915 & 1.26 & 1.498 & 1.969 & 2.515 \\ 
		11 & 0.645 & 0.901 & 1.243 & 1.479 & 1.943 & 2.483 \\ 
		13 & 0.641 & 0.891 & 1.232 & 1.465 & 1.925 & 2.460 \\ 
		15 & 0.638 & 0.883 & 1.223 & 1.455 & 1.912 & 2.443 \\ 
		17 & 0.635 & 0.877 & 1.216 & 1.447 & 1.902 & 2.43 \\ 
		19 & 0.633 & 0.872 & 1.211 & 1.441 & 1.894 & 2.419 \\ 
		21 & 0.631 & 0.868 & 1.207 & 1.436 & 1.887 & 2.411 \\ 
		23 & 0.629 & 0.865 & 1.203 & 1.432 & 1.882 & 2.404 \\ 
		25 & 0.627 & 0.862 & 1.200 & 1.428 & 1.877 & 2.398 \\ 
		27 & 0.626 & 0.860 & 1.197 & 1.425 & 1.873 & 2.393 \\ 
		29 & 0.625 & 0.858 & 1.195 & 1.423 & 1.869 & 2.388 \\ 
		35 & 0.621 & 0.853 & 1.190 & 1.416 & 1.861 & 2.378 \\ 
		45 & 0.616 & 0.848 & 1.184 & 1.410 & 1.853 & 2.367 \\ 
		55 & 0.613 & 0.845 & 1.181 & 1.406 & 1.847 & 2.360 \\ 
		65 & 0.611 & 0.843 & 1.178 & 1.403 & 1.843 & 2.355 \\ 
		75 & 0.608 & 0.841 & 1.176 & 1.400 & 1.840 & 2.351 \\ 
		85 & 0.607 & 0.840 & 1.175 & 1.399 & 1.838 & 2.348 \\ 
		95 & 0.605 & 0.839 & 1.174 & 1.397 & 1.836 & 2.346 \\ 
		$\infty$ & 0.593 & 0.831 & 1.164 & 1.386 & 1.821 & 2.327 \\ 
		\multicolumn{7}{p{4.5 in}}{%
			$N$ is the total number of observations, including the observation for which $Q_{\rm E}$ is calculated. Quantiles for odd $N$ were calculated from eq. \eqref{eqn:5} or (for the asymptotic values in the last row) eq. \eqref{eqn:22}, and are rounded to three decimal places. The probabilities are lower tail probabilities, so that quantiles are upper critical values for significance $1-p$. To use the table for intermediate values of $N$, use the next lower value of $N$ in the relevant part of the table (odd or even). See also Appendix 2 for discussion of interpolation methods for $N \gg 95$.}
	\end{tabular}
\end{table}

\section{Multiple comparisons among IID data}

So far we have provided probabilities or critical values for a single observation from a data set. In most studies, however, it is likely that researchers will calculate a value for $Q_{\rm E}$ for all value/uncertainty pairs in a given set of data to aid in locating the less obvious anomalies (often near-central values with small reported uncertainties). Critical values for a single observation are then too low, potentially leading to a high false discovery rate. 

The most common solution to this problem is to adjust \textit{p}-values or, if calculating quantiles, adjust the probability, for the number of comparisons. Most such adjustment methods assume that the various observations are independent. For a statistic constructed from pairwise comparisons, this is no longer the case; the calculated values of the statistic for each data point depend to some extent on all of the remaining observations. To assist in the analysis of multiple values, therefore, we provide additional tables constructed from extensive simulation. These tables, provided as Table \ref{table:table3}, are constructed so that, given the null hypothesis of equal locations for all data, a proportion \textit{p} of data sets will include one or more values of $Q_{\rm E}$ above the critical value given. (Details of the simulations used to construct  Table \ref{table:table3} are given in Appendix 2).

\begin{table}
	\caption{\label{table:table3}Multiple-observation upper quantiles for median scaled absolute difference}
	\subfloat[$N$ even]{%
		\begin{tabular}{p{0.2in}p{0.5in}p{0.5in}p{0.5in}} 
			$N$ & \multicolumn{3}{p{1.4in}}{Probability $p$} \\ 
			& 0.95 & 0.99 & 0.999 \\ 
			4 & 2.100 & 2.566 & 3.119 \\ 
			6 & 2.104 & 2.520 & 3.022 \\ 
			8 & 2.118 & 2.509 & 2.985 \\ 
			10 & 2.135 & 2.511 & 2.971 \\ 
			12 & 2.153 & 2.518 & 2.968 \\ 
			14 & 2.170 & 2.527 & 2.969 \\ 
			16 & 2.187 & 2.537 & 2.972 \\ 
			18 & 2.202 & 2.548 & 2.976 \\ 
			20 & 2.216 & 2.558 & 2.982 \\ 
			22 & 2.230 & 2.567 & 2.987 \\ 
			24 & 2.242 & 2.577 & 2.993 \\ 
			26 & 2.254 & 2.586 & 2.999 \\ 
			28 & 2.266 & 2.594 & 3.005 \\ 
			30 & 2.276 & 2.603 & 3.011 \\ 
			40 & 2.322 & 2.640 & 3.039 \\ 
			50 & 2.358 & 2.670 & 3.063 \\ 
			60 & 2.389 & 2.696 & 3.085 \\ 
			70 & 2.415 & 2.718 & 3.103 \\ 
			80 & 2.438 & 2.738 & 3.119 \\ 
			90 & 2.458 & 2.756 & 3.134 \\ 
			100 & 2.476 & 2.772 & 3.149 \\ 
		\end{tabular}
	}
	\quad
	\subfloat[$N$ odd]{%
		\begin{tabular}{p{0.2in}p{0.5in}p{0.5in}p{0.5in}} 
			$N$ & \multicolumn{3}{p{1.4in}}{Probability $p$} \\ 
			& 0.95 & 0.99 & 0.999 \\ 
			3 & 2.030 & 2.523 & 3.100 \\ 
			5 & 2.065 & 2.488 & 2.997 \\ 
			7 & 2.089 & 2.484 & 2.964 \\ 
			9 & 2.112 & 2.491 & 2.954 \\ 
			11 & 2.134 & 2.501 & 2.953 \\ 
			13 & 2.154 & 2.513 & 2.956 \\ 
			15 & 2.173 & 2.525 & 2.96 \\ 
			17 & 2.190 & 2.537 & 2.966 \\ 
			19 & 2.205 & 2.548 & 2.973 \\ 
			21 & 2.22 & 2.558 & 2.979 \\ 
			23 & 2.233 & 2.569 & 2.986 \\ 
			25 & 2.246 & 2.578 & 2.993 \\ 
			27 & 2.258 & 2.587 & 2.999 \\ 
			29 & 2.269 & 2.596 & 3.006 \\ 
			35 & 2.299 & 2.620 & 3.024 \\ 
			45 & 2.340 & 2.655 & 3.051 \\ 
			55 & 2.374 & 2.683 & 3.074 \\ 
			65 & 2.402 & 2.707 & 3.094 \\ 
			75 & 2.426 & 2.728 & 3.111 \\ 
			85 & 2.447 & 2.747 & 3.127 \\ 
			95 & 2.467 & 2.764 & 3.141 \\ 
		\end{tabular} 
	}\\

		$N$ is the total number of observations, including the observation for which a given $Q_{\rm E}$ is calculated. Quantiles are provided such that for normally distributed IID data, a proportion \textit{p} of data sets will contain no values greater than the quantiles given. The tables are accordingly suitable for use in outlier testing when the median scaled absolute difference is calculated for all data points in a set of size $N$ and the largest are examined. Quantiles for both odd and even N were estimated from 100 runs, each using $\min(2\times 10^{6}, 10^{7}/N)$ data sets (at least $10^{7}$ quantile determinations in all), followed by smoothing using an appropriate polynomial regression of mean quantile on $\log{n})$. The residual standard deviation for all regressions was typically $10^{-3}$ or less. To use the table for intermediate values of $n$, use the next lower value of $n$ in the relevant part of the table (odd or even). 
	
\end{table}

\begin{figure}
	\centering
	\sidesubfloat[]{%
		\label{fig:4a}
		\resizebox{3.5 in}{!}{%
			\includegraphics[trim={0 0 0 0.25in }, keepaspectratio=true]{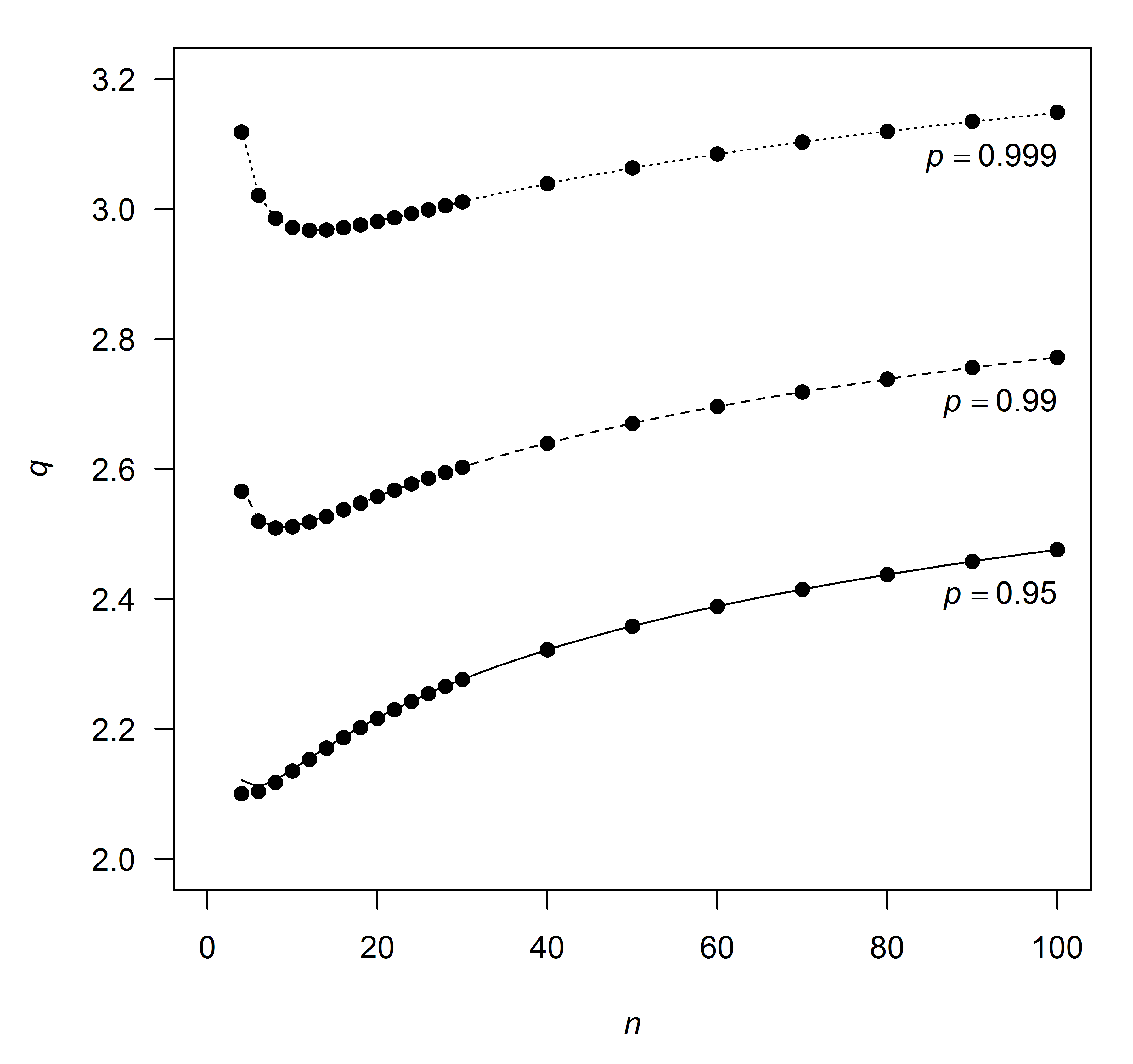} } 
	}
	
	\sidesubfloat[]{%
		\label{fig:4b}
		\resizebox{3.5 in}{!}{%
			\includegraphics[trim={0 0 0 0.25in }, keepaspectratio=true]{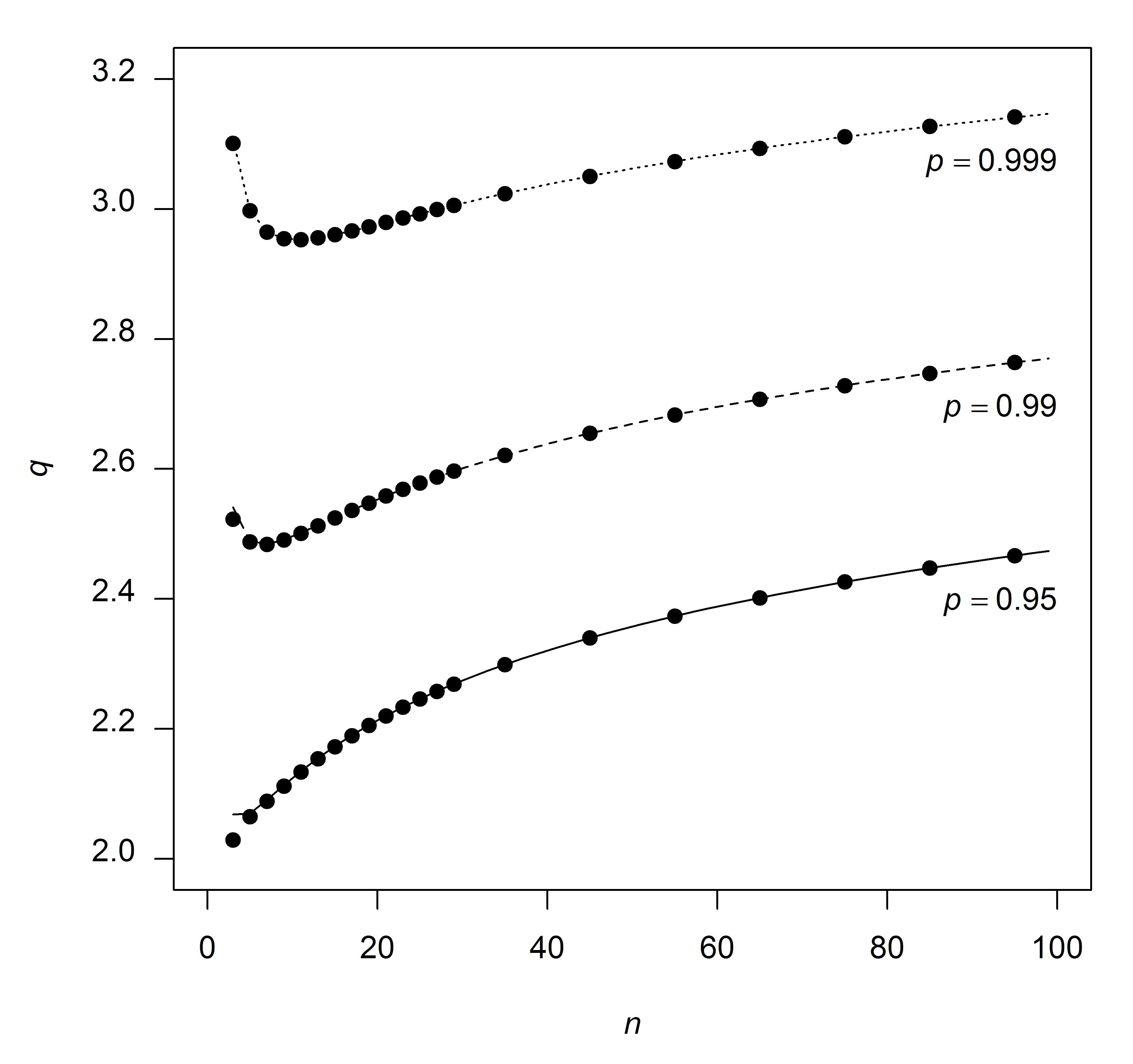} } 
	}
	\caption{Multiple-comparison quantiles from Table \ref{table:table3} (solid circles) for probability $p=0.95$, 0.99, and 0.999 compared with quantiles calculated directly from equations \eqref{eqn:3} and \eqref{eqn:5} (lines) using a probability of $p^{1/n}$. a) Even $n$; b) odd $n$.}
	\label{fig:multquant}
\end{figure}

Perhaps surprisingly, however, the simulations used to construct Table \ref{table:table3} also suggest that lack of independence does not present much of a problem for simple \textit{p}-value adjustment. Figure~\ref{fig:multquant} compares quantiles from Table \ref{table:table3}, obtained from extensive simulation, with quantiles calculated using equations \eqref{eqn:3} and \eqref{eqn:5} using probabilities adjusted for multiple observations assuming independence. The adjustment is simply 
$$
	p_{1} =p_{n}^{1/n} 
$$ 
in which $p_{1}$ is the probability for a single observation falling below the given quantile (as used in equations \eqref{eqn:3} and \eqref{eqn:5}) and $p_{n}$ the probability that all of a set of $n$ values fall below the same quantile. The agreement is very good, with deviations visible only for $n<6$ and even then inconsequentially small. This appears to follow from the relatively small correlations among different values for $Q_{\rm E}$ in the same data set, which fall off quite rapidly with increasing $n$. We conclude that for most practical purposes ordinary \textit{p}-value adjustment methods (for example, Holm \cite{Holm1979} or Benjamini-Hochberg ) should provide reasonably safe inference for multiple comparisons, provided that researchers remember that even with this relatively outlier-resistant indicator, one high value in a set will still tend to increase $Q_{\rm E}$ slightly for other observations.

It is worth considering which multiple comparison strategy is appropriate for typical applications. For a single laboratory considering its own result, single-observation quantiles (\ref{table:table2}) should be appropriate. For an outlier test on a complete data set, where the intent is to limit the probability of falsely detecting an outlier when the true means are equal, Holm correction of p-values or reference to \ref{table:table3} would be appropriate. However, very often in examining interlaboratory study data, the researcher is considering which laboratory or laboratories in a data set should be further examined or marked as suspect, when it is quite likely that there are at least some laboratories meriting closer inspection. For this intermediate purpose it is important to balance detection power against the risk of identifying too many laboratory results as suspect. False discovery rate control as provided by Benjamini-Hochberg adjustment \cite{BH1995} and similar methods is therefore likely to provide a more useful balance of power and protection against over-identification of individual laboratories for general inspection purposes.   

\floatsetup[figure]{style=plain,subcapbesideposition=top}

\section{Critical values and significance levels for heteroscedastic data}

\begin{figure}
	\centering
	\sidesubfloat[]{%
		\label{fig:5a}
		\resizebox{3.2 in}{!}{%
			\includegraphics[trim={0 0 0 0.25in }, keepaspectratio=true]{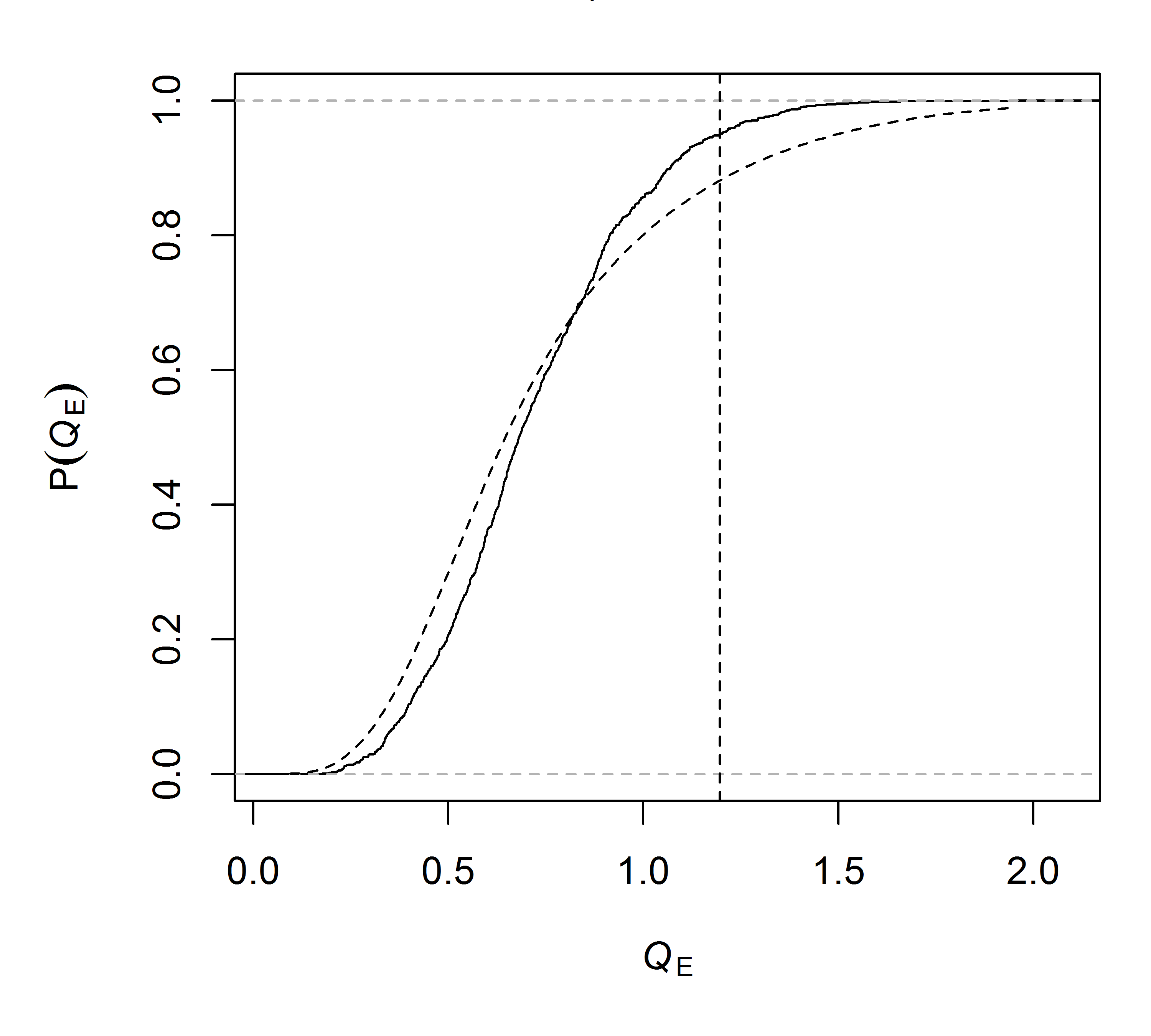} } 
	}
	
	\sidesubfloat[]{%
		\label{fig:5b}
		\resizebox{3.2 in}{!}{%
			\includegraphics[trim={0 0 0 0.25in }, keepaspectratio=true]{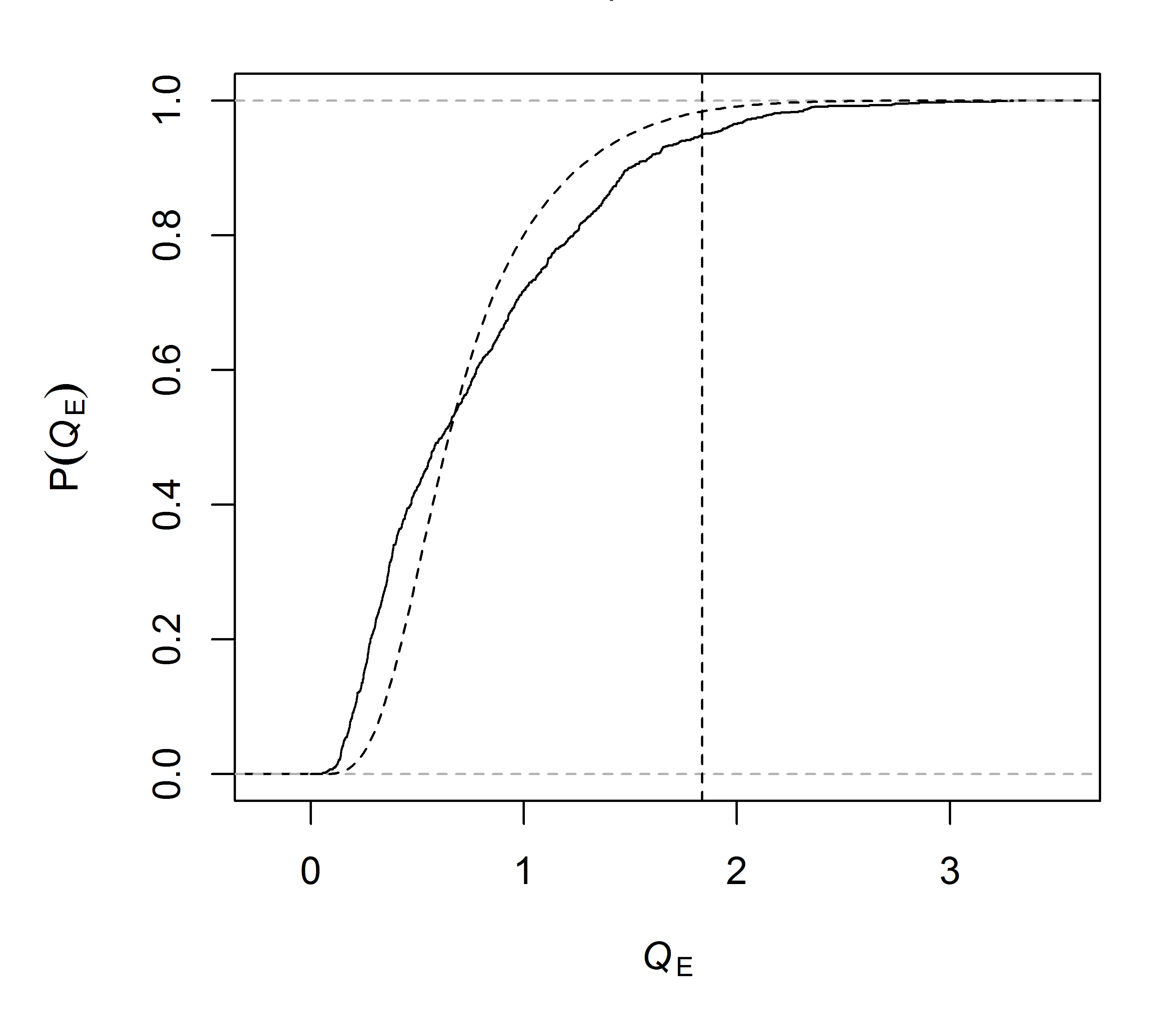} } 
	}
	\caption{Empirical cumulative distributions (solid, stepped curve) of values of $Q_{\rm E}$ in 1000 simulated studies of 10 laboratories, all with the same mean value, nine with standard uncertainty of 1.0 and one, for which the distribution is shown, with standard uncertainty of a) 0.3 and b) 3.0. The cumulative distribution shown as a dashed line is the theoretical distribution assuming homoscedasticity; the vertical dashed line is the 0.95 quantile of the simulated values.}
	\label{fig:ecdf-heteroscedastic}
\end{figure}

In practice, the statistic $Q_{\rm E}$ is expected to be most useful for data with appreciably different stated standard errors or standard uncertainties. For such cases, the tabulated critical values in Table \ref{table:table2} and Table \ref{table:table3} are at best a guide. This can be seen most simply by considering two small simulations; one involving a single laboratory with unusually small (but accurately reported) standard uncertainty and one for the case where a single laboratory has an accurately reported high standard uncertainty. Figure~\ref{fig:ecdf-heteroscedastic} shows the two outcomes. For the case with one unusually small standard uncertainty (0.3 against nine others of 1.0), the distribution of $Q_{\rm E}$ is narrower and the upper 0.95 quantile, at about 1.2,  is appreciably below the 95\% critical value of about 1.5 in Table \ref{table:table2}. For the laboratory with unusually large standard uncertainty of 3.0, the distribution is broader, more positively skewed than the theoretical IID curve, and has an upper 0.95 quantile of 1.84, appreciably above the tabulated value. In this small simulation, the small-uncertainty laboratory exceeds the theoretical critical value in less than 1\% of cases, while the large-uncertainty case shows 10\% of the observations as significant at the 95\% level.

\floatsetup[figure]{style=plain,subcapbesideposition=top}

In principle, one might contemplate specific critical values, or (for observed $Q_{\rm E}$), \textit{p}-values, using adaptations of equations \eqref{eqn:3} and \eqref{eqn:5} with one choice of standard deviation associated with one observation and another standard deviation associated with all other observations. In practice, even this is unlikely to be sufficient for accurate probabilities. In a realistic inter-laboratory study, all the reported uncertainties will differ from one another. For practical application with heteroscedastic data, then, we must either adopt a relatively simple `rule of thumb' and rely on careful follow-up inspection, or use computational methods that incorporate the actual reported uncertainties t provide a case-specific indication.

A simple `rule of thumb' guideline can be obtained from the IID case. Inspection of the single-observation quantiles in Table \ref{table:table2} shows that the critical values at the 95\% level do not exceed 1.7 and for 99\% they only reach 2.0 for $n<10$. For the multiple-observation tables, the 95\% quantiles are between 2.0 and 2.5 and the 99\% quantiles are all above 2.5. As a working rule for inspection, therefore, we suggest that a value of $Q_{\rm E}$ above 2 might reasonably be viewed as meriting further inspection. Figure~\ref{fig:ecdf-heteroscedastic} suggests that this guideline should work reasonably well when uncertainties differ by as much as a factor of 3. In modest sized sets of data where some protection against false discovery rate is desired, this threshold can be increased to 2.5 to avoid overly frequent inspection. We checked these guidelines by simulation using variances randomly generated from a $\chi^{2}$ distribution with three degrees of freedom giving uncertainties (as standard deviation) differing by more than a factor of 10; across data sets of size 5 to 25, individual values of $Q_{\rm E}$ above 2.0 appeared with approximately 1\% probability; data sets containing $Q_{\rm E}> 2.5$ appeared with probability from 1\% to 3\% as $n$ increased, broadly in line with the expected IID probabilities.

Of course, a `suspicious' observation is not necessarily `wrong'; suspicion simply implies a need for closer investigation. One immediate check is to ask whether the value in question has a particularly large uncertainty, which (noting Figure~\ref{fig:ecdf-heteroscedastic}) is more likely to generate higher $Q_{\rm E}$ by chance.

Where more rigorous inference is required for heteroscedastic data, a small simulation or parametric bootstrap is the most straightforward method of examining the range of $Q_{\rm E}$ values likely under the hypothesis that all observations arise from populations with the same mean value. Around 2000 iterations is usually sufficient to give approximate 99\% upper tail quantiles and takes a few seconds for the small data set sizes encountered in metrology comparisons. All that is required is to generate B sets of random data with mean zero and the reported standard deviation for each data point, and calculate $Q_{\rm E}$ for the resulting B data sets. Empirical quantiles provide approximate critical values applicable to each data point and a rough but case-specific \textit{p}-value can be obtained by checking the proportion of simulated $Q_{\rm E}$ values above the observed $Q_{\rm E}$ for a particular observation. Normal \textit{p}-value adjustment for multiple comparison (for example, those suggested by Holm \cite{Holm1979} or by Benjamini and Hochberg \cite{BH1995}) provides reasonable protection against inflated false discovery rates when examining a whole data set simultaneously. For zero counts, we suggest retaining an estimated (uncorrected) \textit{p}-value of 1/B and reporting as ``less than 1/B'' to provide a conservative \textit{p}-value distinguishable from nonzero counts. This can then be corrected upwards in the normal way for multiple comparisons to give, again, a relatively conservative non-zero \textit{p}-value.

\section{Median absolute scaled difference as an indicator of anomaly}

\begin{figure}
	\includegraphics*[width=5.00in, keepaspectratio=true]{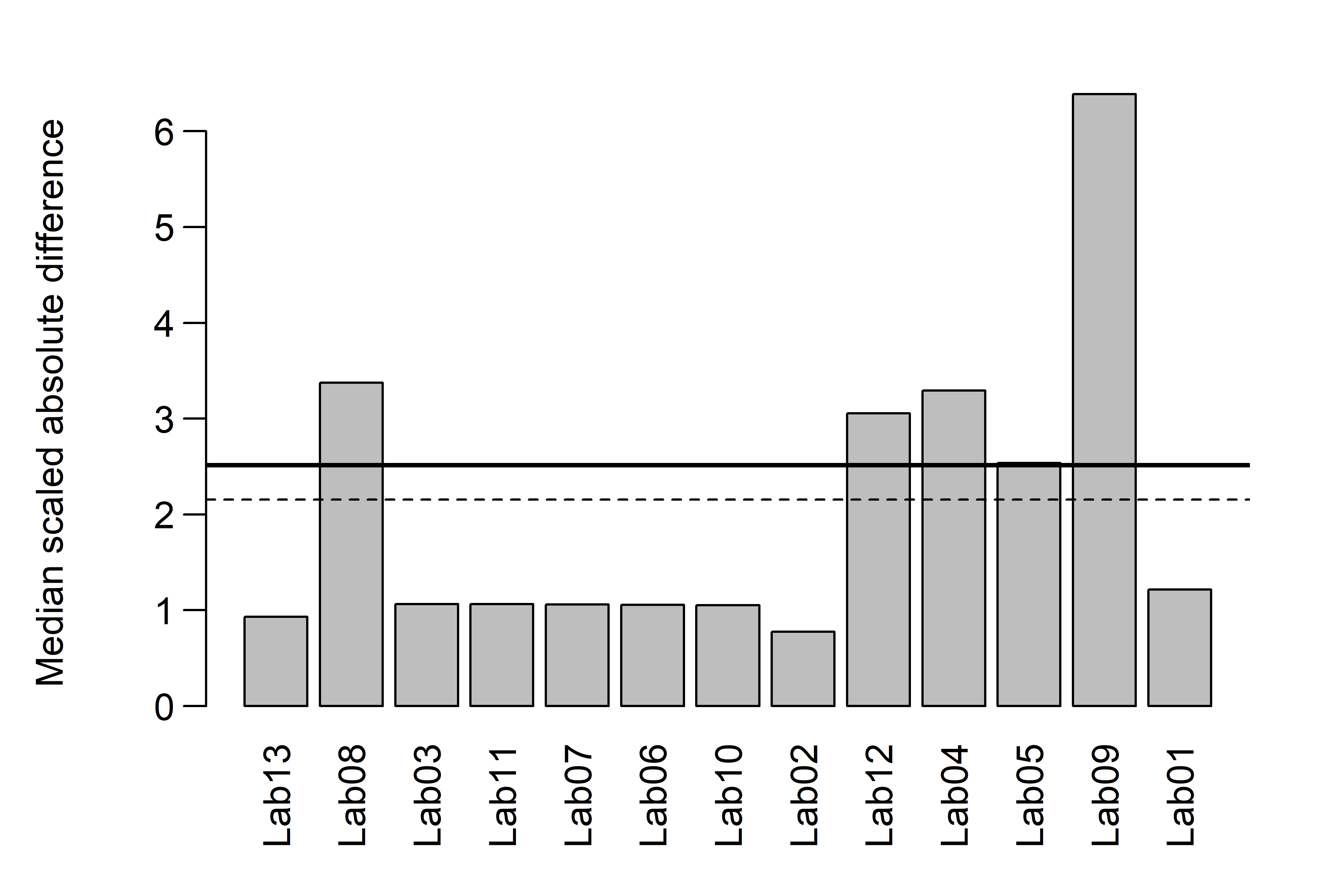}
	\caption{Median absolute scaled difference $Q_{\rm E}$ for the results in Table \ref{table:table1}. Horizontal lines are the multiple-observation 95\% (dashed line) and 99\% (solid line) quantiles for 13 results, taken from Table \ref{table:table3}. The corresponding single-observation quantiles (Table \ref{table:table2}a) would be 1.465 and 1.925 respectively.}
	\label{fig:p23-msd-iid}
\end{figure}

Returning to the data set of Table  \ref{table:table1}, it is now possible to use the MSD indicator to explore whether there are laboratories that show relatively poor agreement with their peers, taking the very different reported uncertainties into account. First, Figure~\ref{fig:p23-msd-iid} shows the calculated MSD values for each result. While Figure~\ref{fig:ccqm-p22} is relatively hard to interpret without the inset, it is immediately clear from inspection of the MSD values that although the range of results may be small, laboratories 4, 8, 9 and 12 show substantial inconsistency with other results when reported uncertainties are taken into account, with laboratory 9 particularly distinct from the majority. Laboratory 5 also shows a need for further consideration. Using the above `rules of thumb', the MSD values for these laboratories are all above 2.0 and all but laboratory 5 substantially exceed the more stringent value of 2.5. For the most extreme results, from laboratories 1 and 13, and the central set of laboratories (Labs 2, 3, 6, 7, 10 and 11) the MSD values are all well below 2.0, giving no immediate reason for concern. Figure~\ref{fig:p23-msd-iid} also shows the multiple-observation 95\% and 99\% quantiles from Table \ref{table:table3}. In this instance, the exact 99\% multiple-observation quantile is very close to 2.5, so the more exact quantiles from Table \ref{table:table3} agree well with the simple rules of thumb. 

\begin{figure}
	\centering
	\includegraphics*[width=5.00in, keepaspectratio=true]{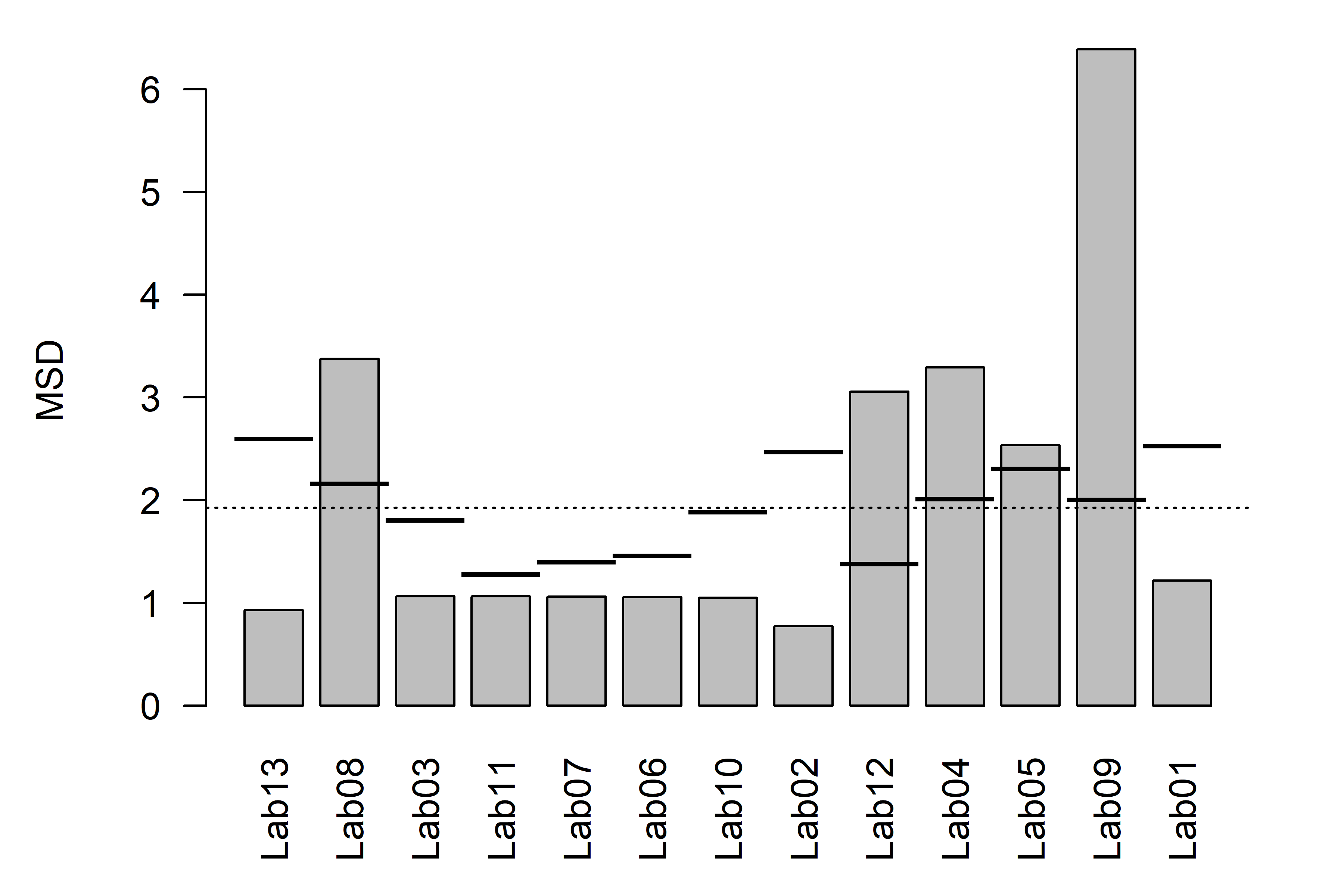}
	\caption{The figure shows the results of a parametric bootstrap for the MSD indicators applied to the results in Table \ref{table:table1}. Short horizontal solid lines across each bar are the laboratory-specific upper 99\% empirical quantiles from a 5000-iteration bootstrap assuming normality. The horizontal dotted line is the 99\% single-observation quantile for 13 results, taken from Table \ref{table:table3}. }
	\label{fig:p23-msd-boot}
\end{figure}

Given that laboratory 5 might be considered marginal, and that some of the apparent inconsistency arises from laboratories with very small reported uncertainties, it is useful to use simulation to examine case-specific quantiles more carefully. Figure~\ref{fig:p23-msd-boot} shows the results from a 5000-iteration bootstrap in the form of laboratory-specific upper 99\% empirical quantiles. The bootstrap tests the null hypothesis that all laboratories have the same mean value and that their results arise from normal distributions with standard deviation equal to the reported standard uncertainty $u$ in Table \ref{table:table1}. For comparison, the single-observation upper 99\% quantile from Table \ref{table:table3} is included. Notice that the laboratories reporting larger uncertainties (Laboratories 1, 2 and 13) show result-specific quantiles higher than the IID quantile from Table \ref{table:table3}, while those reporting smaller uncertainties attract lower (more stringent) case-specific quantiles from the simulation. Despite this, and the fact that Figure~\ref{fig:p23-msd-boot} shows single-observation quantiles, the broad interpretation using the IID quantiles in Figure~\ref{fig:p23-msd-iid} is largely unchanged. 

For a quantitative evaluation, approximate \textit{p}-values can be calculated from the simulation by taking the proportion of simulated results above the observed MSD in each case. Using a Holm correction for 13 comparisons gives $p~<~2.6\times10^{-3}$ for laboratories 4, 8, 9 and 12, with $p=0.005$ for laboratory 5. (The value of $2.6\times10^{-3}$ arises from the suggested procedure above for handling zero observed counts in the simulation). No other results show significant inconsistency, though laboratories 6, 7 and 11 show marginal results (p between 0.05 and 0.10). In this case, then, the case-specific bootstrap results, with Holm correction for multiple comparisons, largely support the immediate interpretation using the multiple-observation quantiles provided by Table \ref{table:table3}, but more strongly suggests that laboratory 5 is also of some concern. 

While this study has largely been chosen to illustrate the use of the MSD indicator, it is worth commenting further on the interpretation. Had one laboratory appeared anomalous, attention would naturally turn to seeking a cause specific to that laboratory. Here, five of thirteen participants appear further from the (largely central) majority than their reported uncertainties would suggest. This is a high proportion. While it is of course useful to seek specific issues among those participants, it is worth noting that a general tendency to underestimate uncertainties across all participants might well give a very similar picture. Given that this was an early pilot comparison, a general tendency to underestimate uncertainties among all participants is perhaps not unlikely, and indeed is not an unusual phenomenon \cite{Thompson2011}. As with most indicators, a high MSD value is a signal that should be followed up, not an immediate proof of individual fault.

\section{Comparison with alternative pairwise indicators}

There is a good deal of literature on identification of outliers in inter-laboratory studies with uncertainty information, and an even more substantial literature on the detection of outlying values in the more general field of meta-analysis. For example, Hedges and Olkin \cite{Hedges1985} provide a thorough treatment of diagnostic procedures for fixed-effect models, and a variety of procedures appropriate to random-effects models has been suggested, including (for example) externally studentized residuals for outlier identification combined with hat values and Cook's distance for identification of influential data points \cite{Viechtbauer2010}. However, most of these approaches rely on a parametric model including a location, with some scaled indicator of distance from, or influence on, the estimate used as an indication of extreme or unusually influential data points. Part of our motivation here is to avoid the need to agree a model for location as a condition for anomaly identification. With this condition, the only comparable indicator so far applied to inter-laboratory studies is the pairwise chi-squared statistic suggested by Douglas and Steele \cite{Steele2006}.  

A complete comparison for a wide variety of circumstances is outside the scope of the present paper. However, it is informative to compare the power and robustness -- in the sense of resistance to secondary anomalies -- of the MSD indicator with that of the pairwise chi-squared indicator. Given that the MSD is based on a median, we might expect somewhat reduced power to detect anomaly compared to the pairwise chi-squared indicator, coupled with higher resistance to secondary anomalies.

\begin{figure} 
	\includegraphics[width=5.00in, keepaspectratio=true]{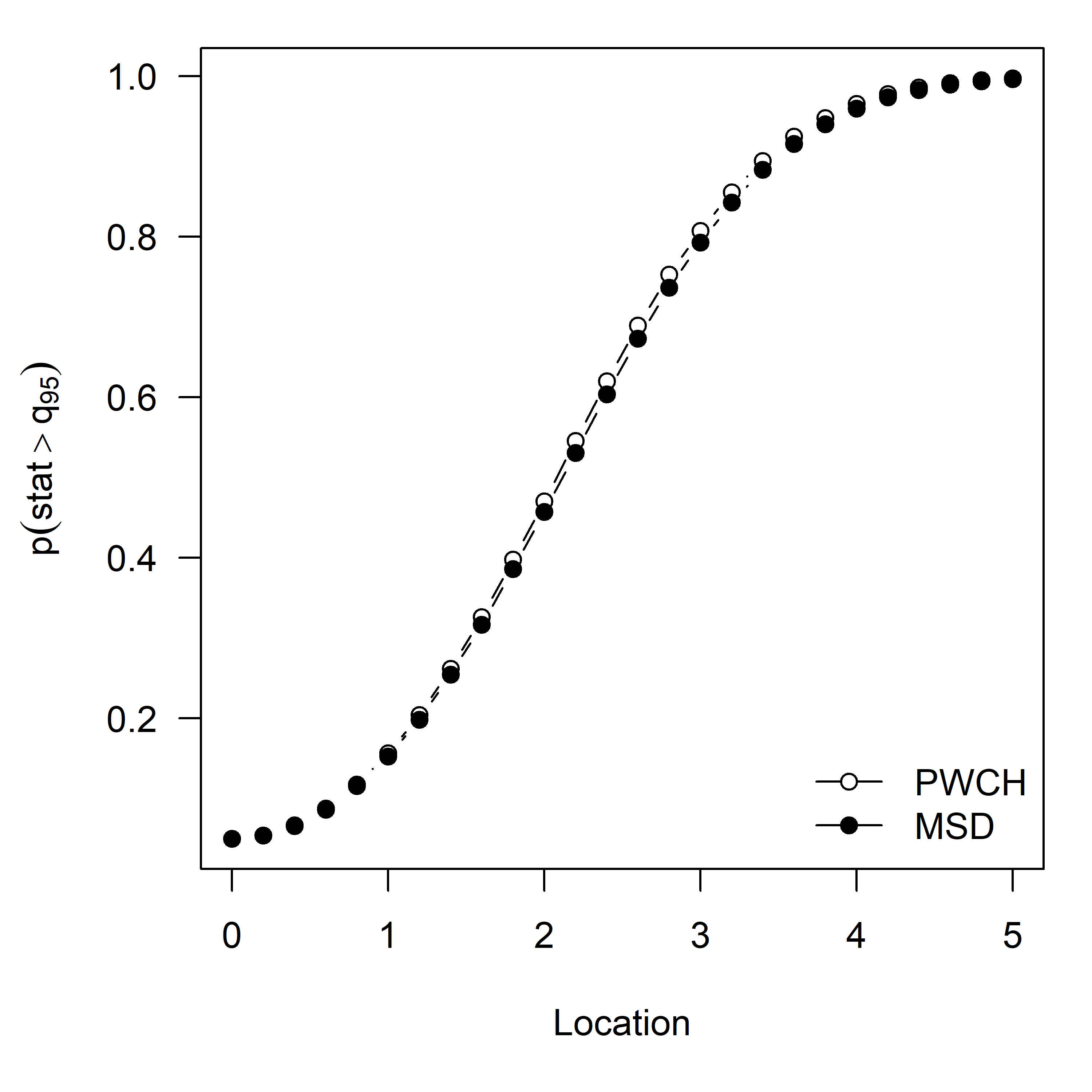}
	\caption{Results of a simulation on 9 data points distributed as $N(0,1)$ and one, for which the corresponding pairwise statistic was calculated, whose location was systematically from 0 to 5. $10^6$ replicates were performed per location. The figure shows the proportion of calculated MSD (``MSD'', solid circles) or pairwise chi-squared values (``PWCH'', open circles) that were above the corresponding upper 0.95 quantile.} 
	\label{fig:power-vs-chisq}
\end{figure}

For the purpose of comparison, two modest simulations were run. Both assumed 10 data points, which is typical of the size of data set for which the indicators are intended. To examine the power at the null, 9 points were set to mean zero, and the location of the tenth, taken to be the laboratory of interest, was varied. 10${}^{6}$ replicates for each point were drawn and the proportion of MSD results for each location that exceeded the 95\% critical value (from Table \ref{table:table2}) was recorded. The same exercise was carried out for the pairwise chi-squared statistic, using a 95\% upper quantile of 2.61, estimated by Monte Carlo simulation using 10${}^{6}$ replicates of a ten-laboratory study with all observations standard normal. The 95\% upper quantile was determined from the whole data set. The resulting comparison of test power is shown in Figure~\ref{fig:power-vs-chisq}. As expected, the power for the MSD is slightly lower than for pairwise chi-squared in the mid-range. However, for this size of data set at least, the degradation in power is clearly inconsequential. 

\begin{figure} 
	\includegraphics[width=5.00in, keepaspectratio=true]{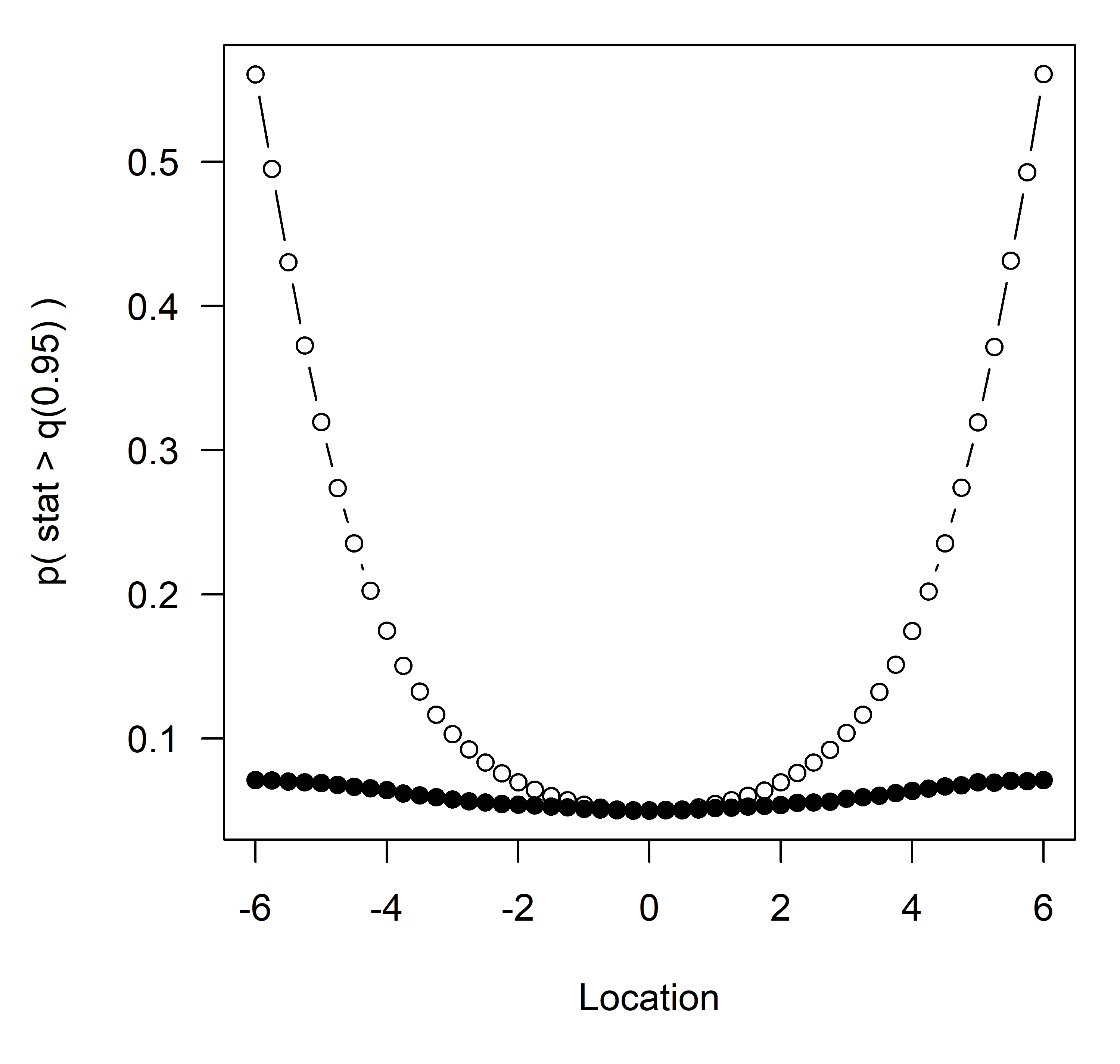}
	\caption {Results of a simulation on 9 data points distributed as $N(0,1)$ and one whose location was systematically from -6 to +6. $10^6$ replicates were performed per location. The figure shows the proportion of calculated MSD (``MSD'', solid circles) or pairwise chi-squared values (``PWCH'', open circles) for one point distributed as $N(0,1)$ that were above the corresponding upper 0.95 quantile.} 
	\label{fig:outlier-resistance}
\end{figure}

The difference in resistance to a second anomalous value was examined using a similar simulation in which the relevant pairwise statistic was calculated for one point with true location at zero, and the location of a second data point varied systematically from -6 to +6. Figure~\ref{fig:outlier-resistance} shows the results. 
Ideally, the proportion of results above the 0.95 quantile (essentially the `false positive rate') would be near 0.05. For MSD, the proportion varies from 0.05 to approximately 0.07 as the `interfering' anomaly becomes more extreme. Further, the false positive rate does not continue to increase once the anomaly is around 6 from the common mean. By comparison, for the pairwise chi-squared statistic, the false positive rate increases rapidly as the outlier passes $\pm$2 continues outwards. The MSD, as intended, accordingly offers substantially greater protection against false positives caused by secondary extreme values. Taking a definition of breakdown point as the proportion of the data set that can move to +$\infty$ before the statistic itself does so, from inspection, the MSD has breakdown of $(N-1)/2N$ (which includes the particular value of interest in $N$).

\section{Conclusions}

We have proposed a new robust pairwise statistic, the pairwise median scaled difference (MSD), for the detection of anomalous location\slash uncertainty pairs in data with associated uncertainties and established its distribution for the IID case. We have shown that it can be used to identify anomalies in typical interlaboratory data and that it offers similar power to a previously provided chi-squared statistic. In addition, the statistic shows very similar power for the identification of an extreme value to the pairwise chi-squared statistic. We have additionally demonstrated the estimation of observation-specific quantiles and \textit{p}-values, using simulation, for a heteroscedastic example. 

Two tables of quantiles have been provided; one that is suitable for checking an individual value in a larger data set, and one that provides a consistent family-wise type 1 error rate. This second table is recommended over the single-observation quantiles for use in identifying anomalous location\slash uncertainty pairs in new data sets, especially as the data set size increases. Using these tables we have shown that, as a simple rule of thumb for identification of suspect values, values of MSD of 2 or more typically merit further investigation. For more exact interpretation, especially where the standard errors or uncertainties differ appreciably, it is recommended that observation-specific quantiles and \textit{p}-values be estimated by simulation, and that the resulting \textit{p}-values be corrected for multiple comparison using, for example, Holm correction or Benjamini-Hochberg false discovery rate control. 

An implementation of the MSD, together with quantile and cumulative probability calculations for the IID case and a parametric bootstrap routine for the heteroscdastic case, is provided in the R metRology package \cite{metRology} under open source license. An interpolation table suitable for rapid calculation of quantiles and probabilities is provided as electronic supplementary information.

Finally, it is additionally worth considering that while the MSD is suggested primarily for the analysis of interlaboratory data, and particularly for CIPM comparisons, the principle should be equally applicable to similar situations in other disciplines, particularly in the inspection of univariate meta-analysis data sets.

\eject


\clearpage

\appendix
\markboth{APPENDICES}{\thepage}

\section{Distribution of the median absolute scaled difference for independent identically distributed observations}
\subsection{The even-$n$ case}

The distribution of $Q_{\rm E}$ for even $n$, which leads to an odd number of differences, is reasonably straightforward for identically distributed data with the same (true) mean and standard deviation $\sigma_{X}$. For $n$ observations, $Q_{\rm E}$ for a single observation chosen at random from the parent distribution is the median of $n-1$ differences drawn randomly from a parent distribution $F(.)$,with density $f(.)$, of the differences.  Let $Y_{1}$ $\mathrm{\le}$ $Y_{2}$ $\mathrm{\le}$ {\dots} $\mathrm{\le}$ $Y_{n-1}$ be the order statistics for those $n-1$ differences. Then the marginal cumulative distribution function of $Y_{\alpha}  (\alpha = 1, 2, {\dots}, n-1)$ is given by
\begin{equation}
F_{Y_{\alpha } (y)} =\sum \limits _{j=\alpha }^{n-1}\left(\begin{array}{c} {n-1} \\ {j} \end{array}\right)\left[F(y)\right]^{j} \left[1-F(y)\right]^{n-j-1}   
\label{eqn:6}
\end{equation}
 
(from, for example, \cite{Mood} p252 ff). The median for $n-1$ observations is $Y_{(}$${}_{n}$${}_{/2)}$ so the cumulative distribution $F_{Q_{{\rm E}} } \left(y\right)$ is given by
\begin{equation}
F_{Q_{{\rm E}} } \left(y\right)=\sum \limits _{j=n/2}^{n-1}\left(\begin{array}{c} {n-1} \\ {j} \end{array}\right)\left[F\left(y\right)\right]^{j} \left[1-F\left(y\right)\right]^{n-j-1}   
\label{eqn:7}
\end{equation}
 
A much more useful form for computation is the incomplete beta function formulation ( \cite{OrderStats}, \cite{Desu1969}):
\begin{equation}
F_{Q_{{\rm E}} } \left(y\right)=I_{F\left(y\right)} \left(r,n-r\right) 
\label{eqn:8}
\end{equation}
 
in which $r=n/2$ and $I_{p}(a, b)$ is the incomplete beta function defined for $a>0$, $b>0$  by
$$
	I_{p} \left(a,b\right)=\frac{1}{B(a,b)} \int \limits _{0}^{p}t^{a-1} (1-t)^{b-1} dt 
$$ 
otherwise known as the cumulative distribution function for the beta distribution with parameters $a$ and $b$. 

It remains to specify the parent distribution $F(y)$ of the scaled absolute differences $D_{i}=|X_i-X_0|\Big/\sqrt{2\sigma_X^2}$, where $x_{0}$ is taken as the observation for the laboratory of interest, $x_{i}$ that for another laboratory $(i = 1, 2, {\dots}, n-1)$ and the square root scales for the variance of a difference of two IID observations. Notice that all of the differences include $x_{0}$, so we cannot simply assume that the differences are independent. Instead, we proceed with a calculation conditional on a value $x_{0}$ for which the differences can be taken as independent draws from a population with $x_{0}$ as a parameter, so that \eqref{eqn:7} applies, with the intention of integrating out $x_{0}$ later. For a given value $x_{0}$ for $x_{0}$, then, the density  $f_{\left|D\right|} \left(\left|d\right||x_{0} \right)$ and cumulative distribution $F_{\left|D\right|} (|d|\, |x_{0} )$ are simply
\begin{equation}
f_{\left|D\right|} \left(\left|d\right||x_{0} \right)=\varphi \left(x_{0} -\left|d_{i} \right|\sqrt{2} \right)+\varphi \left(x_{0} +\left|d_{i} \right|\sqrt{2} \right) 
\label{eqn:9}
\end{equation}
 
and
\begin{equation}
F_{\left|D\right|} \left(\left|d\right||x_{0} \right)=\Phi \left(x_{0} +\left|d_{i} \right|\sqrt{2} \right)-\Phi \left(x_{0} -\left|d_{i} \right|\sqrt{2} \right) 
\label{eqn:10}
\end{equation}
 
Substituting the conditional probabilities into equation \eqref{eqn:7} gives
\begin{equation}
F_{Q_{{\rm E}} } \left(\left|d\right||x_{0} \right)=\sum \limits _{j=n/2}^{n-1}\left(\begin{array}{c} {n-1} \\ {j} \end{array}\right)\left[F_{\left|D\right|} \left(\left|d\right||x_{0} \right)\right]^{j} \left[1-F_{\left|D\right|} \left(\left|d\right||x_{0} \right)\right]^{n-j-1}   
\label{eqn:11}
\end{equation}
 
Integrating over $x_{0}$  (taking $x_{0}\sim N(0,1)$)
\begin{equation}
F_{Q_{{\rm E}} } \left(\left|d\right|\right)=\sum \limits _{j=n/2}^{n-1}\left(\begin{array}{c} {n-1} \\ {j} \end{array}\right)\int \limits _{-\infty }^{\infty }\left[F_{\left|D\right|} \left(\left|d\right||x_{0} \right)\right]^{j} \left[1-F_{\left|D\right|} \left(\left|d\right||x_{0} \right)\right]^{n-j-1} \varphi (x_{0} )dx_{0}    
\label{eqn:12}
\end{equation}
 
or, in the incomplete beta function form,
\begin{equation}
F_{Q_{{\rm E}} } \left(\left|d\right|\right)=\int \limits _{-\infty }^{\infty }I_{F_{\left|D\right|} \left(\left|d\right||x_{0} \right)} \left(r,n-r\right)\varphi (x_{0} )dx_{0}   
\label{eqn:13}
\end{equation}

\subsection{The odd-n case}

Let $x_{1}$, $x_{2}$, {\dots}., $x_{n}$${}_{-1}$ be a random sample from a density f(.) with cumulative distribution F(.). We use n-1 here because we intend to apply the results to the median absolute scaled difference calculated for one observation among n.  Let $Y_{1}$~$\mathrm{\le}$~$Y_{2}$~$\mathrm{\le}$ {\dots}.$\mathrm{\le}$~$Y_{n}$${}_{-1}$ denote the corresponding order statistics. Standard results give the joint density of any two of these as
\begin{equation}
\begin{array}{l} {f_{Y_{\alpha } ,Y_{\beta } } (x,y)=\frac{(n-1)!}{(\alpha -1)!(\beta -\alpha -1)!(n-\beta -1)!} } \\ {\, \, \, \, \, \, \, \, \, \, \, \, \, \, \, \, \, \, \, \, \, \, \, \, \, \, \, \, \times \left[F(x)\right]^{\alpha -1} \left[F(y)-F(x)\right]^{\beta -\alpha -1} \left[1-F(y)\right]^{n-\beta -1} f(x)f(y)I_{(x,\infty )} (y)} \end{array} 
\label{eqn:14}
\end{equation}
 
For a median of $n-1$ observations with odd $n$, the two order statistics of interest have $\alpha = (n-1)/2$ and $\beta = \alpha+1$. Letting $j=(n-1)/2$ and noting that $n-j-1=j$, the above joint density reduces to
\begin{equation}
f_{Y_{\alpha } ,Y_{\beta } } (x,y)=\frac{(n-1)!}{(j-1)!(j)!} \left[F(x)\right]^{j-1} \left[1-F(y)\right]^{j} f(x)f(y)I_{(x,\infty )} (y) 
\label{eqn:15}
\end{equation}
 
Noting that the binomial coefficient $\left(\begin{array}{c} {n-1} \\ {j} \end{array}\right)=\frac{(n-1)!}{j!j!} $, this can also be written 
\begin{equation}
f_{Y_{\alpha } ,Y_{\beta } } (x,y)=j\left(\begin{array}{c} {n-1} \\ {j} \end{array}\right)\left[F(x)\right]^{j-1} \left[1-F(y)\right]^{j} f(x)f(y)I_{(x,\infty )} (y) 
\label{eqn:16}
\end{equation}
 
As above, for the scaled absolute differences $D_{i}=\left|X_{i} - X_{0} \right|\big/ \sqrt{2\sigma _{X}^{2} } $, where $x_{0}$ is taken as the observation for the laboratory of interest, $x_{i}$ that for another laboratory $(i = 1, 2, {\dots}, n-1)$, and the square root scales for the variance of a difference of two IID observations, we replace $f(.)$ and $F(.)$ with the density $f_{\left|D\right|} \left(\left|d\right||x_{0} \right)$ and cumulative distribution $F_{\left|X-x_{0} \right|} (|d|\, |x_{0} )$, conditional on a given value $x_{0}$ for $x_{0}$, described respectively by
\begin{equation}
f_{\left|D\right|} \left(\left|d\right||x_{0} \right)=\varphi \left(x_{0} -\left|d\right|\sqrt{2} \right)+\varphi \left(x_{0} +\left|d\right|\sqrt{2} \right) 
\label{eqn:17}
\end{equation}
 
and
\begin{equation}
F_{\left|D\right|} \left(\left|d\right||x_{0} \right)=\Phi \left(x_{0} +\left|d\right|\sqrt{2} \right)-\Phi \left(x_{0} -\left|d\right|\sqrt{2} \right) 
\label{eqn:18}
\end{equation}

and $f(y)$ and $F(y)$ are replaced by corresponding expressions conditional on $Y_{0}$=$Y_{0}$. 

Cumulative probabilities can be found for an `even' median (odd $n$ in our case) using the cumulative probability for a median $M$ in random samples of size $n-1 = 2r_{e}$;
\begin{equation}
F_{M} (m)=\frac{2}{B(r_{{\rm e}} ,r_{{\rm e}} )} \int \limits _{-\infty }^{m}\left[F(x)\right]^{r-1} \left\{\left[1-F(x)\right]^{r} -\left[1-F(2m-x)\right]^{r} \right\}f(x)dx  
\label{eqn:19}
\end{equation}

where $B(a, b)$ is the beta function (see, for example, \cite{Desu1969} or \cite{OrderStats}). On substitution of $F_{\left|D\right|} \left(\left|d\right||x_{0} \right)$ and noting that this is not defined below zero, this leads to:
\begin{equation}
F_{Q_{x} } \left(\left|d\right||x_{0} \right)=\frac{2}{B(r_{{\rm e}} ,r_{{\rm e}} )} \int \limits _{0}^{\left|d\right|}\left[F_{\left|D\right|} \left(t|x_{0} \right)\right]^{r-1} \left\{\left[1-F_{\left|D\right|} \left(t|x_{0} \right)\right]^{r} -\left[1-F\left(2\left|d\right|-t|x_{0} \right)\right]^{r} \right\}f_{\left|D\right|} \left(t\right)d(t)  
\label{eqn:20}
\end{equation}
 
requiring integration over both the arbitrary integration variable $t$ and over $x_{0}$ to obtain cumulative probabilities. In practice we found that the numerical integral took a few seconds to calculate a complete cumulative probability curve of around 100 points. Quantiles can be found by root finding (slow but accurate as long as the numerical integral remains stable) or by interpolation to a more than adequate approximation for routine use.

\subsection{Asymptotic (large $n$) distribution}

Using the fact that for very large $n$, the distributions of the median for odd and even $n$ converges, we consider only the even-$n$ case for which (since we have $n-1$ differences) the median absolute difference is a single order statistic. We use the incomplete beta form shown in eq. \eqref{eqn:13}. As $n$ increases, the incomplete beta function $I_{p} \left(a,b\right)$ tends to an indicator function $1_{p\in [a/(a+b),1]} $ defined on [0,1] with value unity on $\left[{a\mathord{\left/ {\vphantom {a \left(a+b\right)}} \right. \kern-\nulldelimiterspace} \left(a+b\right)} ,\, 1\right]$ and zero elsewhere. In our application $a=r= n/2$ and $b = n-r$; for large $n$, ${a\mathord{\left/ {\vphantom {a \left(a+b\right)}} \right. \kern-\nulldelimiterspace} \left(a+b\right)} $ tends to 0.5. \eqref{eqn:13} then becomes
\begin{equation}
\mathop{\lim }\limits_{n\to \infty } F_{Q_{{\rm E}} } \left(\left|d\right|\right)=\int \limits _{-\infty }^{\infty }1_{\left(F_{\left|D\right|} \left(\left|d\right||x_{0} \right)>0.5\right)} \varphi (x_{0} )dx_{0}   
\label{eqn:21}
\end{equation}
 
where we have indicated the dependence on $F_{\left|D\right|} \left(\left|d\right||x_{0} \right)$. Importantly, note that $F_{\left|D\right|} \left(\left|d\right||x_{0} \right)$ for any {\textbar}d{\textbar} decreases as {\textbar}$x_{0}${\textbar} increases, so the integration domain for $x_{0}$ is the region $\left|x_{0} \right|$ from zero up to $\left|x_{0} \right|$ such that $F_{\left|D\right|} \left(\left|d\right||x_{0} \right)=0.5$. Noting that $\left. \left|d\right|\right|x_{0} $ is independent of the sign of $x_{0}$ and $x_{0}$ is centred on zero and integrating over $\phi$($x_{0}$), \eqref{eqn:21} reduces  to 
\begin{equation}
\mathop{\lim }\limits_{n\to \infty } F_{Q_{{\rm E}} } \left(\left|d\right|\right)=2\Phi \left(\left. \left|d\right|\sqrt{2} \, \right|\, F_{\left|D\right|} \left(\left|d\right||x_{0} \right)=0.5\right)-1 
\label{eqn:22}
\end{equation}
 
where $\Phi$ is the cumulative distribution function for the standard normal distribution.

Algorithmically, this requires locating the median value of {\textbar}d{\textbar} in $F_{\left|D\right|} $ (which can be done fast by root finding) and returning the corresponding cumulative probability for the half-normal distribution with standard deviation 2${}^{-0.5}$.  A further application of root finding allows location of quantiles for $\mathop{\lim }\limits_{n\to \infty } F_{Q_{{\rm E}} } \left(\left|d\right|\right)$.

\section{Practical estimation of quantiles for the IID case}

Numerical integration of equation \eqref{eqn:3} is sufficiently fast for modest, even  $n$ to allow single probabilities to be calculated explicitly on demand for the single-observation case, (typically taking under 1 second for 100 such quantiles anywhere up to $n=30$). Numerical stability for the even-$n$ case seemed good up to at least $n=10^{6}$, 4-5 orders of magnitude above any common interlaboratory study size. Numerical root finding to locate quantiles is slower but again sufficiently rapid for small numbers of quantiles. However, for multiple quantiles, calculation of quantiles becomes sufficiently slow to notice, taking seconds per quantile. For odd $n$, integration of eq. \eqref{eqn:5} proved feasible at least up to $n=299$, though we found that numerical rounding starts to become a problem at high probabilities and $n$ near 400, which we attribute to the combination of multiple numerical integration and the large dynamic range of the two terms in the central subtraction for high $n$. It is, however, considerably slower than for the even-$n$ case; too slow and computer-intensive for routine use. For such use, interpolation on precalculated tables offers adequate accuracy and very much faster response. Here, we describe the methods used for precalculation and subsequent interpolation. 

For interpolation tables, probabilities were calculated for quantiles at 49 regularly spaced values of $q/(1+q)$ from 0 to 0.8, together with two additional quantiles,  at $0.674/\sqrt{2}$ and at $q/(1+q)=1.0$ ($q= +\infty$). For even $n$, probabilities were calculated by numerical integration of equation 7, implemented in its beta density form for numerical stability at high $n$, for all $n$ from 4 to 30, every whole ten and intervening 4 (e.g. 80, 84, 90, 94,  ...) to 100, and for $10^{k}$ and $5\times10^{k}$ for $k=2$ to $k=5$. The asymptotic (large $n$) values from eq. \eqref{eqn:22} were appended to the table for infinite $n$. For odd $n$, probabilities were obtained by numerical integration of eq. \eqref{eqn:5}, for all odd $n$ from 3 to 29, then odd multiples of 5 to 95, then at intervals of 20 from 109 to 189. For interpolation of odd values thereafter, even-$n$ probabilities (above) were added to the table. The resulting tables are provided as electronic supplementary material.

For interpolated probabilities for tabulated values of $n$, we recommend use of a monotonic spline such as that of Hyman \cite{Hyman1983}, fitting tabulated probabilities as a function of $q/(1+q)$. This appears to cope well with the relatively sharp increase in cumulative probability near $0.674/\sqrt{2}$ for larger $n$. For values of $n$ that are not tabulated, we suggest cubic interpolation of probabilities at each quantile on $n/(n+1)$ to generate a set of interpolated probabilities for each new $n$, followed by monotonic spline interpolation on $q/(q+1)$ as above. A quadratic interpolation on $n/(n+1)$ suffices where interpolation is between the two final tabulated values of $n/(n+1)$. Comparison of spline interpolation using these methods with quadrature integration at arbitrary test points other than spline knots (also provided as a separate table for validation) showed agreement better than 0.0005 for quantile estimates determined from accurate integration.

For determination of quantiles from probabilities, we recommend numerical root-finding on the probability interpolation spline described above; that is, numerically selecting a quantile such that the returned probability is equal to that required. A possible `inverse' strategy, modelling the tabulated quantiles as a function of calculated probability, was examined briefly but was not found useful. First, the cumulative probability is almost constant (near 1.0) above about $q=2.5$ and transposition of quantile and probability results in near-vertical gradient at the upper extreme (and at the lower end for high $n$), with severe adverse effects on spline fitting and interpolation. Second, transposing the model generates a very different interpolation spline to that for predicting probabilities from quantiles, with the result that probabilities found for a known quantile do not return the starting quantile accurately when using the transposed interpolating function. We accordingly recommend root-finding on the probability interpolation spline as above, rather than refitting the quantiles as a function of probability.

For the multiple-comparison cases, Table \ref{table:table3} gives quantiles estimated by simulation. The comparison in Figure~\ref{fig:multquant} shows that for outlier detection, interpolation on single-value quantiles using simple adjusted probabilities is likely to be sufficient for most practical purposes. We therefore do not provide additional interpolation tables for the multiple-comparison case.

\end{document}